\documentclass[sigconf,authorversion,nonacm]{acmart}

\PassOptionsToPackage{table,xcdraw}{xcolor} 
\usepackage{xcolor} 

\usepackage{dirtytalk} 
\usepackage{wrapfig}
\usepackage{subcaption}
\usepackage[bottom]{footmisc} 
\usepackage{booktabs}
\usepackage{graphicx}
\usepackage{multirow}
\usepackage{pifont}
\usepackage{colortbl} 

\usepackage{algorithm}
\usepackage{algpseudocode}
\usepackage[table,xcdraw]{xcolor}

\usepackage{tikz}  
\usepackage{xparse}

\definecolor{myBlue}{RGB}{0, 122, 255}
\definecolor{myBlack}{RGB}{0, 0, 0}
\definecolor{myRed}{RGB}{190,55,74}

\newcommand{\sayit}[1]{{\textit{\say{#1}}}}

\newcommand{\sysname}{AccessParkCV}

\newcommand{\githubrepo}{\hyperlink{https://github.com/makeabilitylab/AccessParkCV}{https://github.com/makeabilitylab/AccessParkCV}}
\newcommand{\huggingfacepage}{\hyperlink{https://huggingface.co/datasets/makeabilitylab/disabilityparking}{https://huggingface.co/datasets/makeabilitylab/disabilityparking}}

\definecolor{blues1}{RGB}{247,251,255}  
\definecolor{blues2}{RGB}{222,235,247}  
\definecolor{blues3}{RGB}{198,219,239}  
\definecolor{blues4}{RGB}{158,202,225}  
\definecolor{blues5}{RGB}{107,174,214}  
\definecolor{blues6}{RGB}{66,146,198}   
\definecolor{blues7}{RGB}{33,113,181}   
\definecolor{blues8}{RGB}{8,81,156}     
\definecolor{blues9}{RGB}{8,48,107}     

\def\getRGB#1,#2,#3\relax{\def\currentR{#1}\def\currentG{#2}\def\currentB{#3}}
\def\getValue#1,#2\relax{\def\currentValue{#1}}

\newcommand{\cellheatmapcolor}[1]{%
    \def\inputvalue{#1}%
    
    \def\colorpointA{0.0,247,251,255}%
    \def\colorpointB{0.1,234,244,250}%
    \def\colorpointC{0.2,222,235,247}%
    \def\colorpointD{0.3,210,227,243}%
    \def\colorpointE{0.4,198,219,239}%
    \def\colorpointF{0.5,178,211,232}%
    \def\colorpointG{0.6,158,202,225}%
    \def\colorpointH{0.7,107,174,214}%
    \def\colorpointI{0.8,44,123,182}%
    \def\colorpointJ{0.9,8,81,156}%
    \def\colorpointK{1.0,8,48,107}%

    \ifdim \inputvalue pt < 0.1pt
        \expandafter\getValue\colorpointA\relax
        \let\valueA=\currentValue 
        \expandafter\getRGB\colorpointA\relax
        \let\RA=\currentR \let\GA=\currentG \let\BA=\currentB
        \expandafter\getValue\colorpointB\relax
        \let\valueB=\currentValue
        \expandafter\getRGB\colorpointB\relax
        \let\RB=\currentR \let\GB=\currentG \let\BB=\currentB
    \else\ifdim \inputvalue pt < 0.2pt
        \expandafter\getValue\colorpointB\relax
        \let\valueA=\currentValue 
        \expandafter\getRGB\colorpointB\relax
        \let\RA=\currentR \let\GA=\currentG \let\BA=\currentB
        \expandafter\getValue\colorpointC\relax
        \let\valueB=\currentValue
        \expandafter\getRGB\colorpointC\relax
        \let\RB=\currentR \let\GB=\currentG \let\BB=\currentB
    \else\ifdim \inputvalue pt < 0.3pt
        \expandafter\getValue\colorpointC\relax
        \let\valueA=\currentValue 
        \expandafter\getRGB\colorpointC\relax
        \let\RA=\currentR \let\GA=\currentG \let\BA=\currentB
        \expandafter\getValue\colorpointD\relax
        \let\valueB=\currentValue
        \expandafter\getRGB\colorpointD\relax
        \let\RB=\currentR \let\GB=\currentG \let\BB=\currentB
    \else\ifdim \inputvalue pt < 0.4pt
        \expandafter\getValue\colorpointD\relax
        \let\valueA=\currentValue 
        \expandafter\getRGB\colorpointD\relax
        \let\RA=\currentR \let\GA=\currentG \let\BA=\currentB
        \expandafter\getValue\colorpointE\relax
        \let\valueB=\currentValue
        \expandafter\getRGB\colorpointE\relax
        \let\RB=\currentR \let\GB=\currentG \let\BB=\currentB
    \else\ifdim \inputvalue pt < 0.5pt
        \expandafter\getValue\colorpointE\relax
        \let\valueA=\currentValue 
        \expandafter\getRGB\colorpointE\relax
        \let\RA=\currentR \let\GA=\currentG \let\BA=\currentB
        \expandafter\getValue\colorpointF\relax
        \let\valueB=\currentValue
        \expandafter\getRGB\colorpointF\relax
        \let\RB=\currentR \let\GB=\currentG \let\BB=\currentB
    \else\ifdim \inputvalue pt < 0.6pt
        \expandafter\getValue\colorpointF\relax
        \let\valueA=\currentValue 
        \expandafter\getRGB\colorpointF\relax
        \let\RA=\currentR \let\GA=\currentG \let\BA=\currentB
        \expandafter\getValue\colorpointG\relax
        \let\valueB=\currentValue
        \expandafter\getRGB\colorpointG\relax
        \let\RB=\currentR \let\GB=\currentG \let\BB=\currentB
    \else\ifdim \inputvalue pt < 0.7pt
        \expandafter\getValue\colorpointG\relax
        \let\valueA=\currentValue 
        \expandafter\getRGB\colorpointG\relax
        \let\RA=\currentR \let\GA=\currentG \let\BA=\currentB
        \expandafter\getValue\colorpointH\relax
        \let\valueB=\currentValue
        \expandafter\getRGB\colorpointH\relax
        \let\RB=\currentR \let\GB=\currentG \let\BB=\currentB
    \else\ifdim \inputvalue pt < 0.8pt
        \expandafter\getValue\colorpointH\relax
        \let\valueA=\currentValue 
        \expandafter\getRGB\colorpointH\relax
        \let\RA=\currentR \let\GA=\currentG \let\BA=\currentB
        \expandafter\getValue\colorpointI\relax
        \let\valueB=\currentValue
        \expandafter\getRGB\colorpointI\relax
        \let\RB=\currentR \let\GB=\currentG \let\BB=\currentB
    \else\ifdim \inputvalue pt < 0.9pt
        \expandafter\getValue\colorpointI\relax
        \let\valueA=\currentValue 
        \expandafter\getRGB\colorpointI\relax
        \let\RA=\currentR \let\GA=\currentG \let\BA=\currentB
        \expandafter\getValue\colorpointJ\relax
        \let\valueB=\currentValue
        \expandafter\getRGB\colorpointJ\relax
        \let\RB=\currentR \let\GB=\currentG \let\BB=\currentB
    \else 
        \expandafter\getValue\colorpointJ\relax
        \let\valueA=\currentValue 
        \expandafter\getRGB\colorpointJ\relax
        \let\RA=\currentR \let\GA=\currentG \let\BA=\currentB
        \expandafter\getValue\colorpointK\relax
        \let\valueB=\currentValue
        \expandafter\getRGB\colorpointK\relax
        \let\RB=\currentR \let\GB=\currentG \let\BB=\currentB
    \fi\fi\fi\fi\fi\fi\fi\fi\fi

    \ifdim \inputvalue pt < 0.05pt
        \cellcolor[RGB]{247,251,255}\color{black}%
    \else\ifdim \inputvalue pt < 0.15pt
        \cellcolor[RGB]{234,244,250}\color{black}%
    \else\ifdim \inputvalue pt < 0.25pt
        \cellcolor[RGB]{222,235,247}\color{black}%
    \else\ifdim \inputvalue pt < 0.35pt
        \cellcolor[RGB]{210,227,243}\color{black}%
    \else\ifdim \inputvalue pt < 0.45pt
        \cellcolor[RGB]{198,219,239}\color{black}%
    \else\ifdim \inputvalue pt < 0.55pt
        \cellcolor[RGB]{178,211,232}\color{black}%
    \else\ifdim \inputvalue pt < 0.65pt
        \cellcolor[RGB]{158,202,225}\color{black}%
    \else\ifdim \inputvalue pt < 0.75pt
        \cellcolor[RGB]{107,174,214}\color{black}%
    \else\ifdim \inputvalue pt < 0.85pt
        \cellcolor[RGB]{44,123,182}\color{white}%
    \else\ifdim \inputvalue pt < 0.95pt
        \cellcolor[RGB]{8,81,156}\color{white}%
    \else
        \cellcolor[RGB]{8,48,107}\color{white}%
    \fi\fi\fi\fi\fi\fi\fi\fi\fi\fi
}

\AtBeginDocument{%
  }

\copyrightyear{2025}
\acmYear{2025}
\setcopyright{cc}
\setcctype{by}
\acmConference[ASSETS '25]{The 27th International ACM SIGACCESS Conference on Computers and Accessibility}{October 26--29, 2025}{Denver, CO, USA}
\acmBooktitle{The 27th International ACM SIGACCESS Conference on Computers and Accessibility (ASSETS '25), October 26--29, 2025, Denver, CO, USA}
\acmDOI{10.1145/3663547.3746377}
\acmISBN{979-8-4007-0676-9/2025/10}

\begin{document}

\title["Where Can \textit{I} Park?"]{"Where Can \textit{I} Park?" Understanding Human Perspectives and Scalably Detecting Disability Parking from Aerial Imagery}


\author{Jared Hwang}
\orcid{0009-0000-4769-1521}
\affiliation{%
    \institution{Allen School of Computer Science}
  \institution{University of Washington, USA}
  \country{}
}
\email{jaredhwa@cs.washington.edu}

\author{Chu Li}
\orcid{0009-0003-7612-6224}
\affiliation{%
    \institution{Allen School of Computer Science}
  \institution{University of Washington, USA}
  \country{}
}
\email{chuchuli@cs.washington.edu}

\author{Hanbyul Kang}
\orcid{0009-0007-4416-0849}
\affiliation{%
    \institution{Allen School of Computer Science}
  \institution{University of Washington, USA}
  \country{}
}
\email{kahaby@cs.washington.edu}

\author{Maryam Hosseini}
\orcid{0000-0002-4088-810X}
\affiliation{%
    \institution{City and Regional Planning}
  \institution{UC Berkeley, USA}
  \country{}
}
\email{maryamh@berkeley.edu}

\author{Jon E. Froehlich}
\orcid{0000-0001-8291-3353}
\affiliation{%
    \institution{Allen School of Computer Science}
  \institution{University of Washington, USA}
  \country{}
}
\email{jonf@cs.washington.edu}
\renewcommand{\shortauthors}{Hwang et al.}


\begin{abstract}
Accessible parking is critical for people with disabilities (PwDs), allowing equitable access to destinations, independent mobility, and community participation. Despite mandates, there has been no large-scale investigation of the quality or allocation of disability parking in the US nor significant research on PwD perspectives and uses of disability parking. In this paper, we first present a semi-structured interview study with 11 PwDs to advance understanding of disability parking uses, concerns, and relevant technology tools. We find that PwDs often adapt to disability parking challenges according to their personal mobility needs and value reliable, real-time accessibility information. Informed by these findings, we then introduce a new deep learning pipeline, called \textit{\sysname}, and parking dataset for automatically \textit{detecting} disability parking and inferring \textit{quality characteristics} (\textit{e.g.,} width) from orthorectified aerial imagery. We achieve a micro-\textit{F1=}0.89 and demonstrate how our pipeline can support new urban analytics and end-user tools. Together, we contribute new qualitative understandings of disability parking, a novel detection pipeline and open dataset, and design guidelines for future tools.

\end{abstract}


\begin{teaserfigure}
  \includegraphics[width=\textwidth]{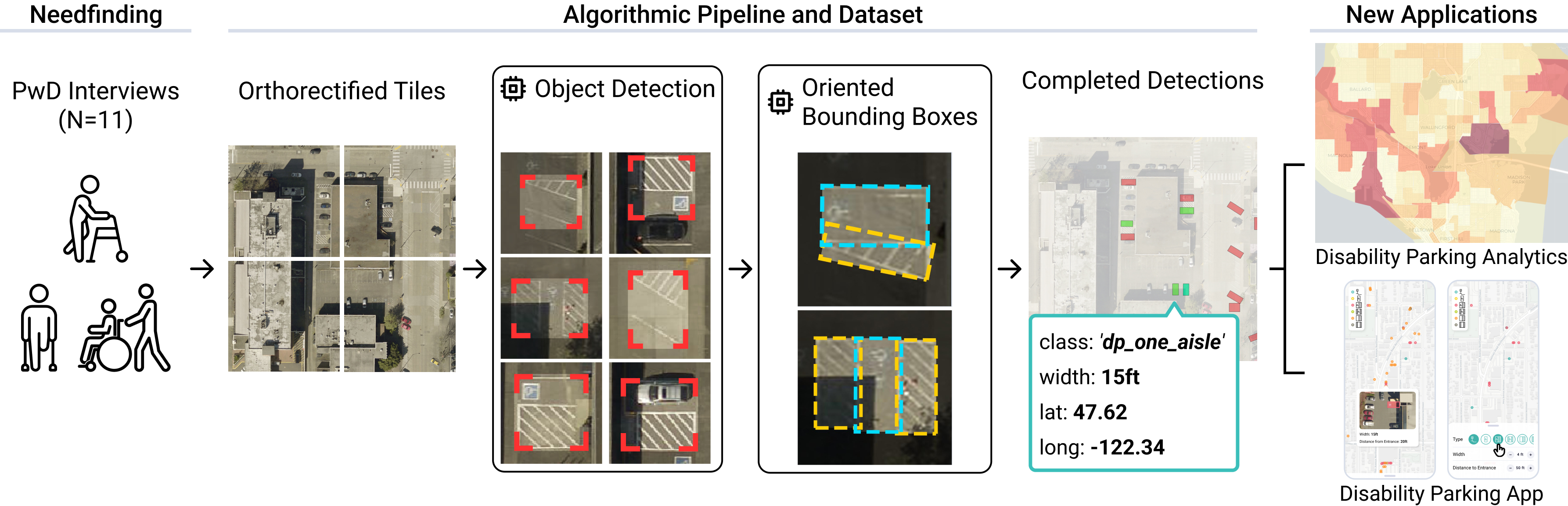}
  \caption{We introduce \textit{\sysname{}}, a novel ML-based disability parking detection and characterization pipeline, composed of an object detection and oriented bounding box model, that (1) locates disability parking from aerial orthoimages and (2) estimates the accessible width of disability parking spaces. We motivate its design using results from a needfinding interview study, and present two case study applications utilizing the pipeline inferences.}
  \Description{The study pipeline, showing needfinding, algorithmic pipeline and dataset, and new applications in sequence. The algorithmic pipeline shows the detection process of disability parking on an aerial image, and the result as the class, width, and latitude longitude of a selected point. New applications show visualizations of the detections as a choropleth map, and a phone app.}
  \label{fig:teaser}
\end{teaserfigure}

\received{16 April 2025}

\maketitle

\section{Introduction}
Accessible parking is critical for people with disabilities (PwDs), allowing equitable access to destinations, independent mobility, and community participation~\cite{ada_standards_for_accessible_design, lee_pwd_transportmode_barriers, laban_dp_privilegeorright}. Although public transit remains a valuable option, it is not always available, equally distributed, or accessible ~\cite{liu_disparities_inpt_usage, saha_projectsidewalk, mwaka_pt_barriers}. Since the \textit{Americans with Disabilities Act} (ADA) of 1990~\cite{ada}, 4-8\% of public parking spaces in the US are required to be reserved for drivers and passengers with disabilities~\cite{laban_dp_privilegeorright}. However, this mandate includes no compliance or verification measures~\cite{ada}, and there has been no systematic large-scale study of the allocation and characteristics (\textit{e.g.}, size) of disability parking in the US. Further, there is limited qualitative research about how PwDs feel about and use disability parking ~\cite{lu_dpparking_width, laban_dp_privilegeorright}.

In this paper, we first present findings from semi-structured interviews with PwDs about their perspectives and uses of disability parking. We then introduce a novel computer vision (CV) pipeline for disability parking inference, called \textit{\sysname}, that not only \textit{identifies} and \textit{locates} disability parking slots from aerial imagery (\textit{e.g.,} plane flyovers) but also \textit{characterizes} their widths. While our algorithms build on emerging work in deep learning to assess built infrastructure from aerial images (\textit{e.g.,} \cite{hosseini_mappingthewalk, ning_sidewalk_2022}), including parking~\cite{ashqer_parkingspace_detection, berry_parkinglot_satellitedetection, yin_satellite_parkingdetection, stopic_satellite_parkingcapacity, drouyer_satellite_parkingoccupancy}, we uniquely contribute new techniques for disability parking and an accompanying labeled dataset for others to build on our work. As mixed-methods research, our research questions (RQs) span both the formative and technical:

\begin{itemize}
    \item \textbf{RQ1}: How do PwDs use disability parking, and how does it fit into their overall transportation profile?
    \item \textbf{RQ2}: How well can CV models locate and characterize disability parking from aerial imagery?
    \item \textbf{RQ3}: How could inferences from such a CV model be used to build technology tools for PwDs and policymakers?
\end{itemize}

To address these questions, we first recruited eleven PwDs for a three-part, semi-structured interview (60-90 minutes). In Part 1, we asked about general transit choice behavior and travel before specifically investigating disability parking in Part 2. In Part 3, we showed thirteen sketches of envisioned technology tools for disability parking, including route planning, \textit{in situ} navigation, and the "out-of-car" experience once parked. From thematic analysis of study transcripts, we find that PwDs' priorities vary significantly with each individual's mobility needs and that such needs are not met by current disability parking policies or implementations. Consequently, PwDs must adapt dynamically when preparing to travel and when parking. We also find that PwDs value real-time and reliable information about accessible parking and places of interest (POIs) in advance of making trips. 

Informed by these qualitative findings, we designed, built, and evaluated a novel disability parking inference pipeline, \sysname, capable of identifying, locating, and characterizing the width of disability parking spots in orthorectified aerial imagery. To train and evaluate our pipeline, we annotated 11,762 parking spots from open-source aerial orthoimages of Seattle, WA~\cite{seattle_dataset}, Washington, DC~\cite{dc_dataset}, and Spring Hill, TN~\cite{springhill_dataset}. Our open-source pipeline achieves state-of-the-art performance: a micro-F1 score~\cite{sokolova_f1score} of 0.89 on detections and an average error of 5.40\% (\textit{SD} = 17.52\%) when estimating parking width. Further, to explore generalizability across datasets and the pipeline's susceptibility to error, we performed two additional investigations: an evaluation on regions outside the training set including at different resolutions and a comparison of the pipeline's parking width error to the error between two human annotators. Moreover, to demonstrate the value of these automated extractions, we introduce two example applications: a geoanalytic disability parking map for policymakers and disability advocates, and an end-user personalized mobile app for seeing and filtering for disability parking.

In summary, we contribute \textbf{formative study findings} about the perceived benefits, uses, and drawbacks of disability parking in the US; a \textbf{new algorithmic pipeline} for detecting and locating disability parking spots from aerial imagery and for characterizing their quality; a \textbf{new disability parking dataset} for others to build on our work; and \textbf{two example applications} demonstrating the potential of our pipeline and dataset. To support open science and enable others to build on our work, both the annotated dataset\footnote{\huggingfacepage} and AccessParkCV code\footnote{\githubrepo} are open source. 

\section{Related Work}

We provide background on the role and implementation of disability parking before contextualizing our work in parking-focused mapping tools and computer vision techniques.

\subsection{Disability Parking}

Disability parking is fundamental to ensuring equitable access and independent mobility for people with disabilities (PwDs)~\cite{lee_pwd_transportmode_barriers, laban_dp_privilegeorright}. In the US, its origins are intertwined with the broader disability rights movement and the push for societal inclusion throughout the latter half of the 20th century~\cite{Shapiro1994NoPity}. Early federal legislation, such as the \textit{Architectural Barriers Act} of 1968~\cite{ArchitecturalBarriersAct1968} and \textit{Section 504} of the \textit{Rehabilitation Act} of 1973~\cite{RehabilitationActSection504_1973}, began mandating accessibility in specific contexts, laying the groundwork for more comprehensive standards. However, it was the passage of the \textit{Americans with Disabilities Act} (ADA) in 1990~\cite{ada} that established enforceable, nationwide requirements for accessible parking in public accommodations and commercial facilities.

The \textit{ADA Standards for Accessible Design}~\cite{ada_standards_for_accessible_design} provide detailed specifications for disability parking spaces. Key requirements include specific minimum dimensions, adjacency to marked access aisles that provide critical space for ramp deployment or maneuvering mobility devices, and clear signage featuring the \textit{International Symbol of Accessibility}. Van-accessible spaces mandate even greater width and vertical clearance, along with specific signage, to accommodate larger vehicles often equipped with lifts~\cite{ada_parking_spaces_guidelines}. The required number of accessible spaces is calculated based on the total parking capacity of a lot, ensuring proportional availability~\cite{ada_standards_for_accessible_design, laban_dp_privilegeorright}.

Despite these federal mandates, manual audits conducted across regions consistently reveal significant deficiencies in the implementation and maintenance of disability parking~\cite{stone_youtakemyspace_itakeyourair, jackson_parking_areusersvoicesheard}. For instance, a survey of 50 public accommodations in Maryland found that 26\% failed to provide the legally required number of accessible spaces, and 32\% lacked a necessary van-accessible space ~\cite{stone_youtakemyspace_itakeyourair}. Common defects included improperly marked or absent access aisles and inadequate accessible routes connecting parking to building entrances~\cite{stone_youtakemyspace_itakeyourair}. Similarly, fieldwork at 36 locations in Melbourne, Australia---which has similar disability parking requirements to the US~\cite{aus_disabilityparking}--- found that none fully complied with recommended best practices, citing issues like undersized spaces, poor line marking, and missing curb ramps~\cite{jackson_parking_areusersvoicesheard}.

These infrastructural shortcomings directly impact the lived experiences and mobility of PwDs ~\cite{jackson_parking_areusersvoicesheard, lu_formalized_dp_system, ross_families_child_disability}. For example, the aforementioned Melbourne study also surveyed 71 disability parking permit holders, revealing widespread feelings of insufficient accessible parking availability, particularly in busy activity centers, along with concerns about the safety of space placement relative to traffic~\cite{jackson_parking_areusersvoicesheard}. Research in Saga, Japan, observed decreased vacancy rates after introducing a formal permit system, with subsequent surveys indicating user frustration regarding lack of availability~\cite{lu_formalized_dp_system}. Furthermore, ethnographic work with families managing childhood disability in Ontario, Canada, demonstrated how parking facilities could be "technically" accessible according to regulations but functionally inaccessible due to factors like timing, poor snow clearance, or temporary obstructions. This often forced families to develop their own coping strategies with little institutional support, highlighting the difference between compliance and true usability~\cite{ross_families_child_disability}. 

Finally, a large body of work investigates disability parking abuse or violations, including motivations~\cite{laban_dp_privilegeorright} and potential enforcement or prevention strategies~\cite{lu_dp_signagestrategies, shoup_ending_dpabuse, eleftheriou_preventing_violations}. Though our research does not focus on violations, such misuse exacerbates the availability issues stemming from inadequate provision or poor design~\cite{shoup_ending_dpabuse}.

In sum, prior research establishes the critical importance of disability parking, details the design standards intended to ensure its utility, and documents common failures in implementation along with the negative consequences for users. Building on this foundation, our work seeks to deepen our understanding of how disability parking affects the planning and travel behavior of PwDs through qualitative inquiry. Further, we introduce a novel disability parking detection pipeline, moving beyond the limitations of localized manual audits highlighted in previous studies.

\subsection{Accessibility and Parking Technology Tools}

Many technology tools aim to provide PwDs a gateway to information about accessibility and infrastructure features in the physical world, including sidewalks~\cite{saha_projectsidewalk}, crosswalks~\cite{ahmetovic_crosswalkdetection}, and POI access~\cite{google_accessiblemaps}---though disability parking itself has received less attention. To enable these tools, varying approaches are used to source data about the features in question, including crowdsourcing and specialized hardware and computer vision. We discuss both accessibility crowdsourcing and parking management systems as two of these information gateways.

Crowdsourcing is a commonly used approach to scalably collect information about the built environment, and many tools are built off of crowdsourced data~\cite{saha_projectsidewalk,wemap,axsmap,wheelmap}. A user survey of the wheelchair accessibility crowdsourcing app, \textit{Wemap}~\cite{wemap}, found that people liked the idea of crowdsourced apps and said that they would use them. A non-accessibility focused survey on crowdsourcing found that people want reliable information that can suggest parking location and dimensions~\cite{margreiter_cowdsourcing_invehicle}. For accessibility specifically, tools such as \textit{AXS Map}~\cite{axsmap}, \textit{EasyWheel}~\cite{easywheel}, \textit{WheelMap}~\cite{wheelmap}, and \textit{SmartBFA}~\cite{smartbfa} are all crowdsourcing platforms that allow users to manually (or passively, in the case of SmartBFA) tag the accessibility of locations, with both AXS Map and EasyWheel including parking information. Although crowdsourcing is a powerful tool, it can suffer from data reliability and requires a user pool to maintain and update accessibility tags~\cite{saha_projectsidewalk, brady_crowdsourcingapps}. Our contribution approaches the problem from a large-scale auditing perspective that does not depend on users to characterize accessible spaces, though crowdsourcing could be used for human review and quality control.

Another avenue for information access to the real-world is parking management systems that provide both real-time information and may help mitigate disability parking violations~\cite{disassist_iot, sarkar_drone_licenseplates, gining_dp_smartparking}. Proposed solutions range from sensors placed on the car or space~\cite{disassist_iot, fikri_smartparking_iot, gining_dp_smartparking, carnelli_parkus_parkingdetection} to camera and machine learning-based parking monitoring~\cite{lee_dpsystem_deeplearning, tschentscher_videobased_parkingspace_detection}. While such systems may allow the user to see availability in a particular lot ahead of time, installing sensors requires hardware and does not address parking that was not initially integrated in the parking management system. Moreover, our focus is on new automated techniques to identify disability parking spaces at scale (\textit{e.g.,} to enable new types of urban analytics), something bespoke infrastructure-sensing solutions cannot achieve.

\begin{figure*}[t]
    \centering
    \includegraphics[width=1\linewidth]{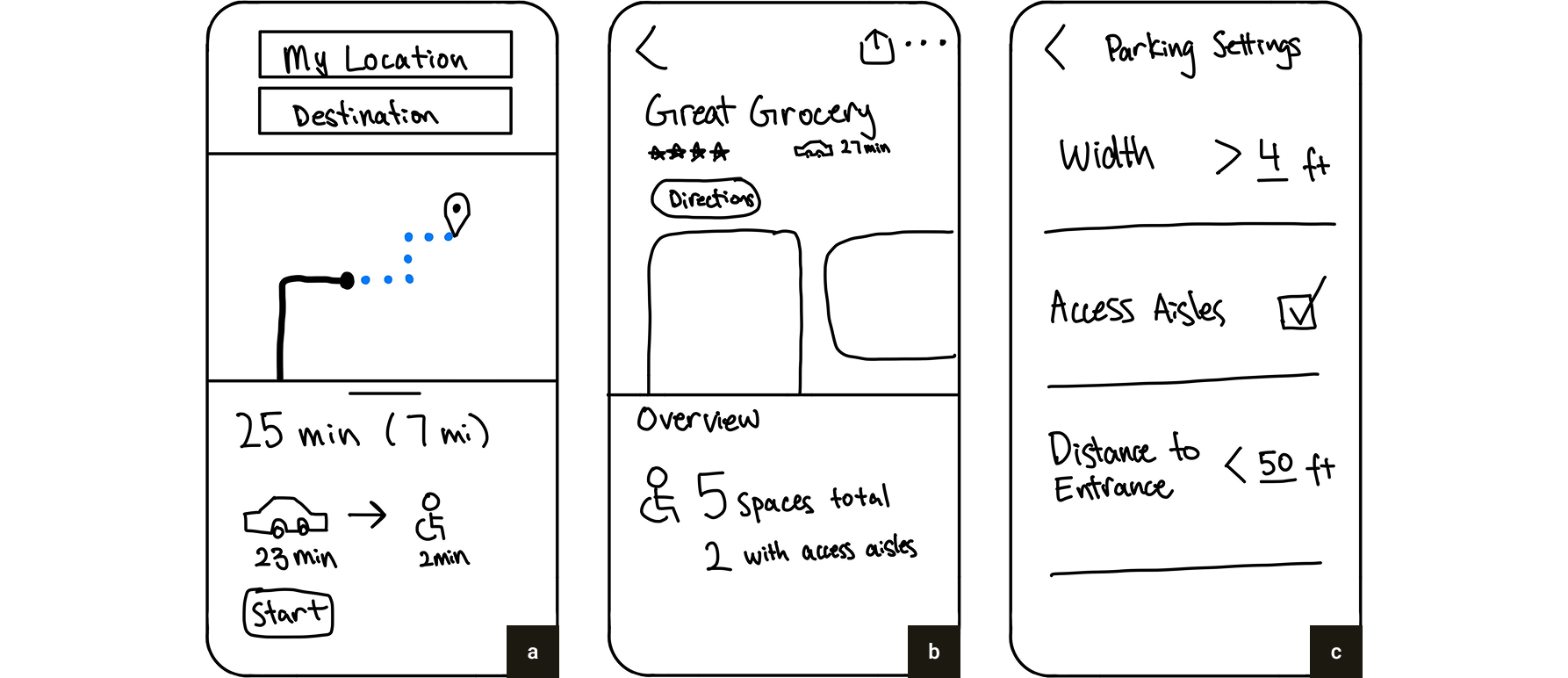}
    \caption{Three example design probes of the 13 total shared with participants: (a) a routing tool showing the path from a parking space to the entrance of a destination, estimating both driving time and time to traverse to the entrance, (b) information about a specific place of interest, including disability parking space counts , and (c) a settings page, where one can filter for disability parking based on personalized needs. See the supplementary materials for the full 13 probes.}
    \Description{Three app interface sketches. The left shows a routing app showing estimated distance and path from parking space to entrance. The middle shows information on how many accessible spaces a grocery store has, and how many with access aisles. The right shows a settings page, to customize the width, access aisles, and distance to entrance desired.}
    \label{fig:design-probes}
\end{figure*}

\subsection{Automatically Detecting Parking Spots}

Our work introduces a novel CV pipeline for disability parking detection, building on literature in automated analyses of the built environment from aerial imagery~\cite{hosseini_mappingthewalk, larkin_builtenvironment_gsv_satellite, albert_cnn_satelliteurbanenvironments, huerta_greenspaces_satellite, kranjcic_greeninfra_satellite, neupane_dl_urbansemanticsegmentation_satllite}. Given the many potential applications of automated parking detection in general---from autonomous driving to parking management---there is a wide array of approaches for detecting parking spaces and occupancy (\textit{e.g.},~\cite{drouyer_satellite_parkingoccupancy},~\cite{huang_inclinedboundingboxes_parking},~\cite{berry_parkinglot_satellitedetection}). However, no such work is dedicated to detecting and characterizing \textit{disability parking} specifically, so we discuss general methods here.

Live occupancy attempts to detect if there is a vehicle in a given space at a given time~\cite{amato_deeplearning_parkingoccupancy}. A common approach involves a static video camera overseeing the parking lot, with parking space positions annotated \textit{a priori} by a human; subsequently, an algorithm attempts to detect the presence of a vehicle within each designated space, including background subtraction~\cite{bibi_parkingdetection_backgroundsubtraction, drouyer_satellite_parkingoccupancy, yamada_vehicleparkingdetection}, feature matching from drone footage~\cite{regester_drone_cv_parkingspaces}, and variations of convolutional neural nets (CNNs)~\cite{nyambal_cnn_parkingspacedetection, amato_deeplearning_parkingoccupancy}. Similar approaches are also used without having a static feed;  \cite{grbic_selfsupervised_parkingdetect} and  \cite{sarkar_drone_licenseplates} determine occupancy via self-supervised learning and drones to examine license plates with CNNs, respectively, without requiring a permanently installed camera. Although these approaches are generally effective at detecting if a car is present, they are not as relevant to our work given that (1) the spaces must already be identified, (2) we do not attempt to determine if a space is occupied in real time, and (3) we make use of aerial, bird's-eye imagery \textit{vs.} infrastructure-mounted cameras.

Another common approach, particularly for automated urban analyses like zone use, is to detect entire parking lots from aerial images, including producing segmentation masks with variations on CNNs (\textit{e.g.}, ~\cite{berry_parkinglot_satellitedetection}, ~\cite{henry_parkinglot_detection_aerial}), or just classifying if an image contains a parking lot~\cite{cisek_transfer_2017}. However, we do not attempt to detect full parking lots, only individual spaces given an arbitrary location.

More relevant to our work are methods to detect if there \textit{is} a parking space and, if so, geometric boundaries. One approach uses aerial imagery and breaks parking lots into their constituent structures  (\textit{e.g.}, lines and boxes), then counts the boxes~\cite{koutaki_parkinglotstructure_fromaerial, seo_hierarchical_parkinglot_aerial, ashqer_parkingspace_detection}---similarly, our characterization model also leverages the rectilinear structure of parking spaces. Other approaches to detecting disability parking use machine learning; Varghese \textit{et al.}~\cite{varghese_detecting_lot_vacancy} use support vector machines and background subtraction to detect both delimited and non-delimited (no demarcating lines) parking spaces. Seo \textit{et al.}~\cite{seo_hierarchical_parkinglot_aerial} use self-supervised learning with $\sim$70\% accuracy.
Lindblom \textit{et al.}~\cite{lindblom_parkingspacedetection_fromaerial_cnn} use CNNs to detect parking spaces from aerial imagery, but they see a high false positive rate in cases where there are many lines in the street or box-like shapes. Most similarly to our work, Huang \textit{et al.}~\cite{huang_inclinedboundingboxes_parking} attempt to generate inclined, or oriented, bounding boxes around parking spaces from aerial footage (as we do to characterize the width of spaces). They achieve precision and recall in the high-90-percent range, depending on the test dataset. We build on the above work but differ in our focus on detecting and characterizing disability parking specifically, and use of a transformer based architecture for our parking detection model.

\section{Study 1: Interview Study of PwDs}
We first present our semi-structured interview study of how PwDs use, feel about, and struggle with disability parking (Study 1) before introducing \sysname{} and example applications partially informed by those formative findings (Study 2).

\subsection{Methods}
To investigate the effects of disability parking availability, quality, and characteristics on PwD travel planning, behavior, and attitudes, we conducted a remote interview study with 11 PwD participants.

\subsubsection{Participants}

Participants were recruited through word of mouth, social media, and disability organizations. Study advertisements linked to a screening survey that collected demographics, disability information, and parking eligibility duration. We selected 11 adult participants (18+) who either had disability parking eligibility themselves or were primary drivers for eligible individuals, all possessing disability parking placards. Because our study was advertised online by some partners, we used \textit{reCAPTCHA} scores embedded in our Qualtrics screener to filter fraudulent responses. See ~\autoref{tab:participant-table} for detailed participant information.

\begin{table}[b]
\footnotesize 
\begin{tabular}{ll >{\raggedright\arraybackslash}p{2.4cm} >{\raggedright\arraybackslash}p{2.0cm} >{\raggedright\arraybackslash}p{1.6cm}} 
\rowcolor[HTML]{FFFFFF}
\multicolumn{1}{l}{\cellcolor[HTML]{FFFFFF}\textbf{P}} & \textbf{Age} & \textbf{Placard Reason} & \textbf{Mobility Aid} & \textbf{Travel Freq.} \\ \hline
\rowcolor[HTML]{FFFFFF}
1 & 25-34 & Temp. Disability & Crutches & > 4/week \\
\rowcolor[HTML]{F3F3F3}
2 & 45-64 & Perm. Disability & Wheelchair & > 4/week \\
\rowcolor[HTML]{FFFFFF}
3 & 45-64 & Perm. Disability & Wheelchair & > 4/week \\
\rowcolor[HTML]{F3F3F3}
4 & 25-34 & Perm. Disability & Wheelchair, Leg Braces, Crutches & > 4/week \\
\rowcolor[HTML]{FFFFFF}
5 & 18-24 & Perm. Disability & Wheelchair & > 4/week \\
\rowcolor[HTML]{F3F3F3}
6 & 45-64 & Primary driver for PwD & Wheelchair & > 4/week \\
\rowcolor[HTML]{FFFFFF}
7 & 45-64 & Perm. Disability & Wheelchair & > 4/week \\
\rowcolor[HTML]{F3F3F3}
8 & 45-64 & Perm. Disability & Wheelchair & > 4/week \\
\rowcolor[HTML]{FFFFFF}
9 & 45-64 & Perm. Disability & Wheelchair & 2-4/week \\
\rowcolor[HTML]{F3F3F3}
10 & 45-64 & Perm. Disability & Wheelchair & > 4/week \\
\rowcolor[HTML]{FFFFFF}
11 & 45-64 & Perm. Disability & Wheelchair & > 4/week
\end{tabular}
\caption{Participant demographics. Travel frequency includes commuting and personal trips.}
\label{tab:participant-table}
\end{table}

\subsubsection{Procedure}

Our semi-structured interview consisted of three sections: 
(1) broad exploration of how PwDs use transportation and disability parking to identify transit choice patterns; 
(2) examination of disability parking experiences, including advantages, disadvantages, and potential improvements; 
and (3) assessment of participants' technology usage habits and feedback on 13 design probes representing potential disability-parking related tools across five thematic areas: route planning, place-of-interest investigation, in situ navigation, out-of-car experiences, and curbside assistance---see examples in ~\autoref{fig:design-probes}. Informed by \cite{hara_assistivetechnologies}, design probes were shown as PowerPoint slides.

Prior to the main study, two researchers conducted a pilot interview with one participant (P1), after which minor changes were made to the interview script to ensure clarity. For the subsequent sessions, two researchers led five interviews each. All interviews were conducted via Zoom, with each session lasting 60-90 minutes. Participants were compensated US\$25 per hour.

\subsubsection{Analysis}
We recorded and transcribed all interview sessions. For analysis, we used a combination of deductive and inductive coding~\cite{braun_thematicanalysis}. To start, we created an initial codebook based on the interview protocol. Two primary researchers then engaged in an iterative process of coding and peer checking~\cite{creswell2016qualitative,LincolnGuba1985, morse_critical_2015} to ensure the reliability and validity of the analysis. 
Of the 11 interviews, one researcher coded five and another coded six. The two researchers then met to discuss and resolve any disagreements, updating the codebook as necessary. The researchers then swapped their set of assigned interviews and spot-checked each others' coding.

\subsection{Results}
Overall, we found that PwDs approach and use disability parking in diverse and personalized ways---often as a function of their mobility disability, prior experiences, and vehicle type. Still, common patterns emerged around preferred disability parking designs, concerns and problems, and opportunities for improvement. We organize our findings around two key research questions: (1) How do PwDs use disability parking? (2) What kind of technology tools or information would PwDs find useful?

\subsubsection{Experiences with Disability Parking}

\begin{figure}[h]
    \includegraphics[width=1\linewidth]{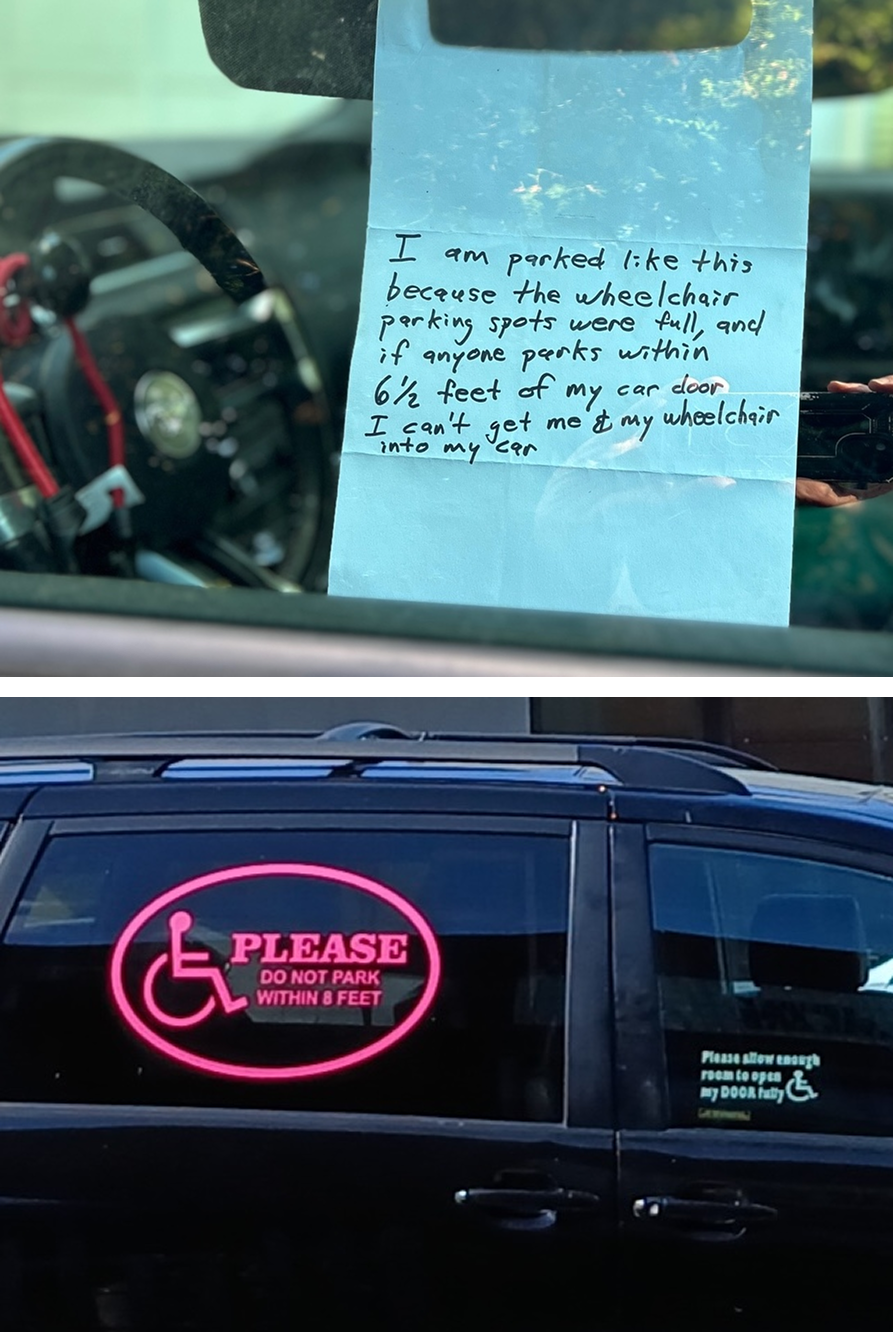}
    \caption{Participants placed decals, signs, and cones on or near their vehicles to warn other drivers not to obstruct the space needed for them or their wheelchair to enter and exit the vehicle. Photos from participants P7 and P2.}
    \Description{Top, a paper sign affixed to the inside of a car window on the driver's side. Handwritten: "I am parked like this because the wheelchair parking spots were full, and if anyone parks within 6 and a half feet of my car door I can't get me \& my wheelchair into my car." Bottom, a van with two decals on the passenger side windows. One shows the international symbol of accessibility and says "Please do not park within 6 feet" in all caps. The other says "please allow enough room to open my door fully."}
    \label{fig:fig-interview-photos}
\end{figure}
Accessible parking is a complex reality for PwDs, shaped by individual mobility, the need for self-devised solutions, and frustrating encounters with poorly designed built environments and public misuse. While regulations exist, their inconsistent implementation creates access barriers. 

\textbf{Personal mobility needs shape parking preferences.} PwDs assess parking spaces based on a range of factors from mobility needs to vehicle size and door locations. P7 noted, for example, that \sayit{if the car can fit in and I can get in and out of the car, that's all I look for.} Interestingly, the need to find accessible parking dominates, even compared to traversal distance to the final destination. P6 observed, \sayit{Let planners know it doesn't have to be really close to the entrance, right? It doesn't. That's not so important.} P9 echoed this but also emphasized the need for accessible routes from the parking location to the destination: \sayit{As long as I can park someplace and use my equipment, I don't mind going further, but other times without sidewalks and curb cuts, there's been times where... I couldn't park anywhere.} Others mentioned the design of spaces themselves: For example, P2 emphasized that the directional location of the aisle is also important, saying \sayit{Most are on the right side, why not put them on the left side? Doesn't make any sense to me, except for if somebody's in a manual chair.} These diverse perspectives emphasize the subjectivity of a preferred disability parking space. Still, P8, who works as a disability advocate, stated that \sayit{the perfectly designed parking spot is truly an ADA spot, and I don't say that about everything that the ADA does; but in terms of parking, it actually did it right.} This indicates that \textit{improper implementation} of disability parking requirements is a key issue, not the regulation itself.

\textbf{Adaptive strategies.} Given inconsistencies in accessible parking, PwDs develop a variety of strategies to ensure their specific requirements are met. A common approach involves relying on the assistance of others, as exemplified by P2, who shared: \sayit{[my husband] would drop me off in front and go park and pick me up.} Similarly, P4 mentioned \sayit{I have two brothers... who have a cute little minivan because of me}, and P6 serves as the primary driver for her son. Beyond human assistance, multiple participants carried a sign or a parking cone to delineate their space or warn others from parking too close, as illustrated in \autoref{fig:fig-interview-photos}. Some reluctantly resort to parking over multiple spaces, acknowledging its inconvenience and potential to frustrate others: P7 mentioned \sayit{I don't like taking two spots at an angle because that pisses people off}, and P3 echoed \sayit{I'll park out over the white line... and then you come back and you can see people who got really annoyed, and it's like, I'm sorry, I can't help it.} These strategies, while effective, highlight the persistent need for PwDs to take extra measures simply to achieve access.

\textbf{Burden of "wrong" and mental calculus for access.} Moreover, employing these adaptive strategies often creates a sense of doing something "wrong" just to gain access. P2 recounted:  \sayit{There's been quite a few times when I've had to either create my own space or illegally park in a bank that was closed.} P2 described relying on facility staff to \sayit{find a spot and put cones in the spot next to me... but it's a pain in the ass and it really, I wish it wasn't like that} (P11). This tension between access \textit{vs.} effort was a persistent backdrop and stressor influencing travel decisions: \sayit{I do math in my head about...how badly do I need to go to the store today?} (P11). 

\textbf{Public misuse.} Beyond structural and implementation issues, the challenges of disability parking were exacerbated by members of the general public. Nearly every participant expressed frustration at drivers without disability parking placards illegally occupying disabled parking spots, reducing availability. P8 lamented that these spots are \sayit{always full of people just parking there for two minutes}, while P5 noted that \sayit{people park on the loading spaces all the time which blocks me from getting in a car completely.} The problem is further compounded by concerns related to confronting individuals, as P10 starkly emphasized: \sayit{Number one, it's not my job to educate the public about disability parking, and number two, people carry firearms. I don't want to get shot.}

\textbf{Alternative transit.} While public transit can be a promising option---and several participants reported using public transit in combination with other travel forms---similar to prior work~\cite{liu_disparities_inpt_usage, mwaka_pt_barriers}, we found that public services are not always available or problematic in other ways. For example, P7 stated, \sayit{I've never taken public transit in Seattle because it's just impractical... it's unpredictable to take public transit.} In contrast, P8 found success with park-and-ride options, noting \sayit{I've found that we drive to [the park and ride] with lovely accessible spots... and it takes me 30 seconds to take the train and get dropped off.} However, P9 highlighted key limitations: \sayit{[Public transit] only runs 8 to 5 and doesn't go out of city limits. It's so limited that I'm forced to have a vehicle.} P6 reflected on the ease of getting on/off accessible buses but pointed to surrounding challenges: \sayit{Being on the bus and getting on the bus is easy, but there's all the stuff around it} (P6), referring to other factors like distance to the bus stop, uphill climbs, and weather conditions.

\subsubsection{Desired Data and Technology Tool Needs}
When asked about \textit{what} information would help participants find and navigate to accessible spaces as well as opportunities for technology support tools to help (as part of our design probes), participants commonly emphasized detailed and up-to-date information about parking occupancy, location, or existence. 

\textbf{Comprehensive routing.} Participants desired accessibility-aware routing not just to a disability parking space but an accessible route from their car to the entrance of their destination. P4 noted, \sayit{Sometimes you want to go somewhere and have no idea how you're going to get there. That's what I like about routing tools, they may not give you the exact [details], but they give you something to work with.} This was highlighted in participants' responses to our probes (\textit{e.g.}, probes \#2, \#3, and \#5): \sayit{that information [showing three spaces, one van space] is super helpful} (P6), \sayit{seeing [the path from the parking space] is helpful, because when I get out of the car, I don't necessarily know whether there's a flight of stairs or the accessible ramp is that way... just having that extra information, it can't not help} (P8); \sayit{oh my gosh, these would be so incredibly helpful. It's kind of... knowing what you're up for} (P10). 

\textbf{Value of dynamic data.} Despite the existence of some current tools that provide static information (\textit{e.g.,} Google Maps shows accessible entrances, seating, and restrooms \cite{google_accessiblemaps}, though often incomplete), participants emphasized that dynamic, real-time data was paramount. P3 articulated this need clearly: \sayit{If you could say 'All these are full; the next closest ADA spot is here,' that would be pretty cool}. Similarly, P6 expressed a desire to \sayit{to know how full the parking lot is;} with P8 echoing, \sayit{that parking garage says it's full, but there's still four accessible spots left. Awesome, I would definitely use an app like that.} Participants also mentioned learning from non-traditional information sources, with one explaining, \sayit{I have received more help from disability influencers who want to take trips than probably any typical brochure} (P4). These desires for accurate, real-time information were particularly visible as positive reception to probes \#3, \#6, \#9, and \#13.

\textbf{Accuracy and trust concerns.} However, there are challenges to real-time information access, both in determining if a parking space exists or if it is full. Some participants expressed skepticism about the reliability of information systems, \sayit{[I don't want] somebody else's computer program getting me the information that I want when I can just do it myself} (P7), in response to probes \#4, \#5, and \#6; \sayit{I don't think I would believe that those numbers are up to date} (P1); \sayit{I don't trust [a tool that shows allocation and location of disability parking]} (P7). The issue of parking violations further complicates the usefulness of information systems since violations decrease the reliability of information about available spaces. As P8 observed, \sayit{The Uber drivers will pull right into the two white stripes because they know they're just going to be there for a minute.}

\begin{figure*}[t]
  \includegraphics[width=1\linewidth]{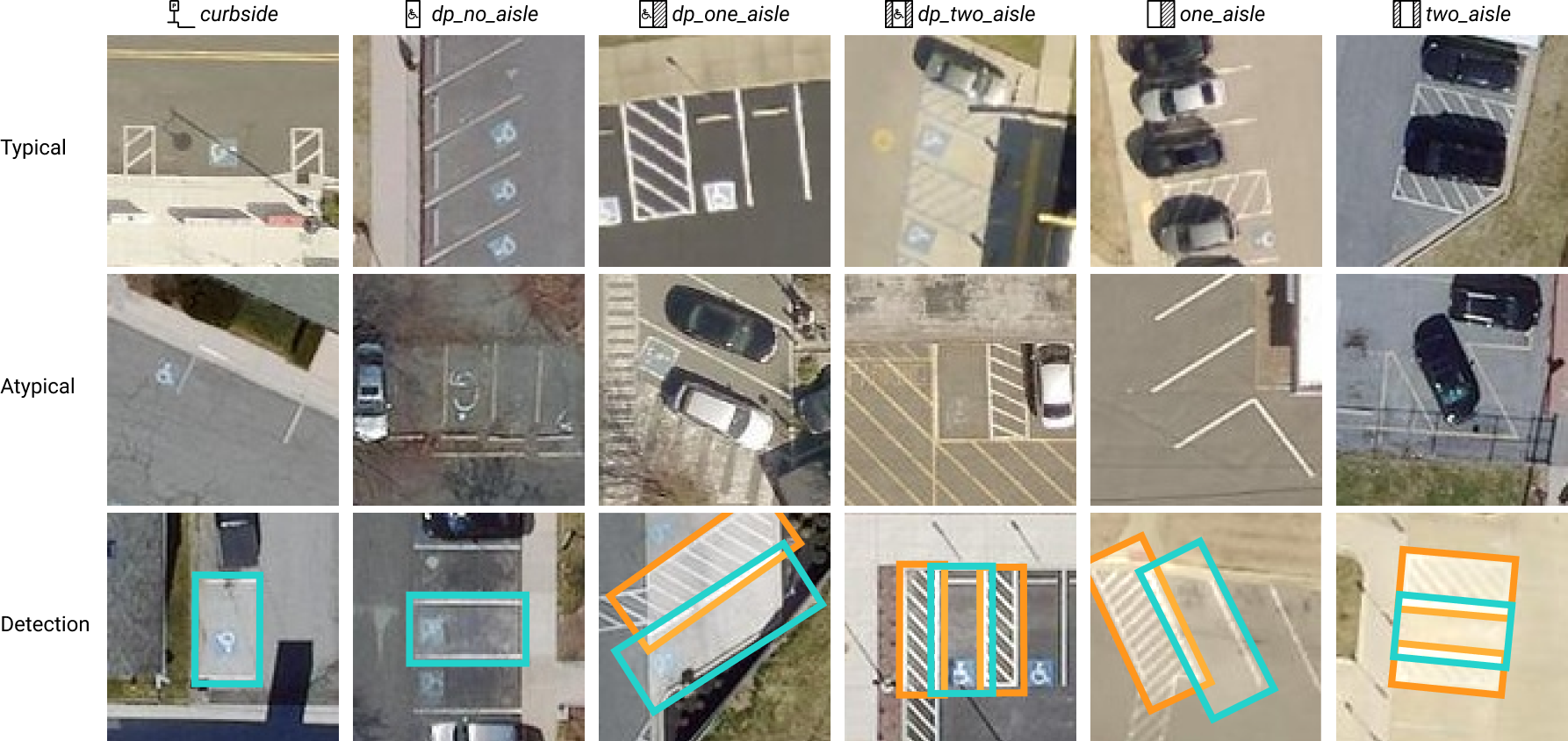}
  \caption{Our dataset contains 11,762 objects in seven categories: \textit{access\_aisle}, \textit{curbside}, \textit{dp\_no\_aisle}, \textit{dp\_one\_aisle}, \textit{dp\_two\_aisle}, \textit{one\_aisle}, and \textit{two\_aisle}. The images show examples for each parking category, with \textit{access\_aisle} implicitly included in the latter four. We include typical and atypical examples, and examples of the resulting detections from our pipeline (with parking detections in cyan and aisle detections in orange).}
  \Description{A grid of images of parking spaces laid out in six columns (curbside, dp no aisle, dp one aisle, dp two aisle, one aisle, two aisle) and three rows (typical, atypical detection). An example of each instance is shown. The bottom row shows colored rectangles overlaid on the area of the parking spaces and aisles.}
  \label{fig:parking_archetypes}
\end{figure*}

\subsubsection{Design Recommendations for Disability Parking.}

Drawing on our formative study and ADA guidelines~\cite{ada_parking_spaces_guidelines}, we synthesize the following design recommendations for disability parking. A key overarching goal of this paper is to develop sensing solutions capable of tracking and ensuring compliance with these recommendations, which \sysname{} in Study 2 partially addresses. 

\begin{enumerate}
  \item\textbf{Spacing:} Provide adequate, ADA mandated~\cite{ada_parking_spaces_guidelines} clearance on one or both sides of the disability parking spot (including at least one access aisle).
  \item \textbf{Signage:} Ensure there is visible signage for each space (as mandated by the ADA~\cite{ada_parking_spaces_guidelines}), indicating disability parking designation and van accessibility. Signage should emphasize that access aisles are \textit{not} for parking and the violation consequences.
  \item \textbf{Maintenance:} Keep parking spaces well maintained, including refreshing fading paint, to ensure that their boundaries and accessible zones are clear and visible.
  \item \textbf{Location:} Avoid placing disability parking on sloped areas or directly adjacent to high traffic zones.
  \item \textbf{Connectivity:} Ensure that all disability parking has an accessible path to the entrance of the destination the lot is intended to serve.
  \item \textbf{Compliance:} Enact policy and auditing steps to ensure that disability parking is allocated according to the ADA required minimums. This is an especially important consideration for public infrastructure projects that remove parking since for some, vehicle travel is the only viable option. 
\end{enumerate}

\section{Study 2: Detecting and Characterizing Disability Parking from Aerial Images}
\label{sec_study2}

Building on our Study 1 findings and established ADA guidelines~\cite{ada_parking_spaces_guidelines}, Study 2 introduces \sysname{}, a novel computer vision (CV) pipeline designed for the scalable identification and characterization of disability parking spaces from aerial imagery. Our pipeline directly addresses the need for robust tools to assess compliance with accessibility standards and to provide actionable data, as highlighted by our participants. In particular, \textit{location}, \textit{spacing}, and \textit{compliance}, from the six design recommendations above; however, our ability to track real-time occupancy or maintenance status is limited given aerial imagery update frequency (\textit{i.e.,} several years between surveys), which we expand on in the Limitations of our work (\autoref{limitations}).

Below, we begin by detailing our dataset and labeling methodology, followed by an explanation of \sysname{}---composed of two key pieces, a \textit{locator} and a \textit{characterizer}---and three empirical evaluations: an overall performance evaluation across test sets from three US regions, an examination of our width characterizer, and an investigation of model performance on regions outside our dataset.

\subsection{Dataset}\label{dataset}

We present and describe our dataset, including orthorectification, the collection and labeling methodology, and the seven inference classes, which allow us to not just detect disability parking spaces but also categorize them into types (\textit{e.g.,} one \textit{vs.} two aisle spaces).

\subsubsection{Orthorectification}

\sysname{} uses aerial images (\textit{e.g.}, from a satellite or a plane flying overhead) with sufficiently high resolution ($\sim$7.6cm/px) that parking slots are visually discernible. These images are typically collected by local or federal governments and released openly (\textit{e.g.,} on ArcGIS Online) but can be distorted due to sensor hardware, terrain structure/elevation, and the curvature of the earth. Orthorectification removes these distortions~\cite{tucker_nasas_2004,zhou_comprehensive_2005}. The orthoimage tile system~\cite{slippytiles} defines each tile by an \texttt{x}, \texttt{y}, and \texttt{z} axis (with \texttt{z} referring to zoom level, a value expressing the scale at which the Earth is rendered into tiles), letting us map each pixel in the image to a precise geospatial coordinate on the Earth's surface.

\begin{figure}[b]
    \centering
    \includegraphics[width=\linewidth]{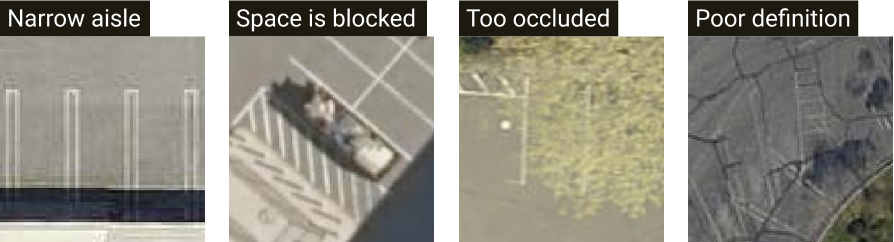}
    \caption{Examples of potential disability parking spaces that we decided \textit{not} to label due to overly narrow aisles, spaces being blocked due to garbage containers or other obstructions, occlusions from trees or shadows, or generally poor definition images.}
    \Description{Four examples of possible parking objects that were not labeled. One has very narrow aisles, about one sixth the width of the parking space. The next shows a large container on top of a parking space with an aisle. The next shows a tree mostly occluding what may be a parking space. The last shows a very blurred and faded parking space that may have an aisle.}
    \label{fig:not-labeled}
\end{figure}

\begin{figure}[b]
    \centering
    \includegraphics[width=1\linewidth]{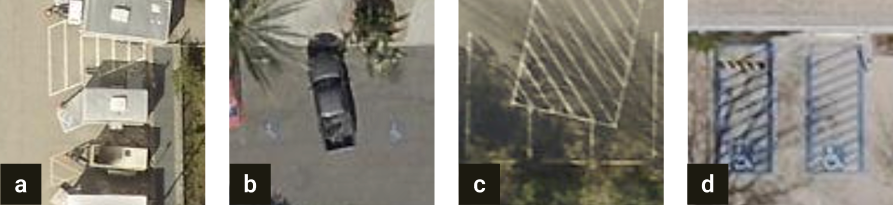}
    \caption{Examples where the research team struggled to determine disability parking and/or their bounds. (a) A space with a logo, which is completely blocked by a temporary structure. (b) A space with non-visible lines. (c) An atypical striped zone painted on top of what may be a parking space. (d) The only case where striping overlaps the logo, creating ambiguity about whether it is a parking space.
    }
    \Description{Four examples of objects that might be parking, but are atypical in certain ways: a large, building like object on top of one parking space, next very faded lines, next a striped zone seemingly randomly placed on top of a parking space, and finally a disability parking space that itself is striped.}
    \label{fig:weird-objects}
\end{figure}

\subsubsection{Data Sampling and Preprocessing}

We source and preprocess the images used for our final labeled dataset in three steps: (1) downloading the aerial imagery as 256×256px tiles from three US cities, (2) stitching the tiles together to 512×512px, and (3) sampling the resultant 512×512 tiles using a small object detection model (separate from our main detection model in \autoref{locating_disability_parking}) to identify candidate parking areas. For the latter, we use a small seed dataset from a fourth city (Denver). We describe each step below. 

In Step 1, we use \textit{Tile2Net}~\cite{hosseini_mappingthewalk} to download our open-source aerial imagery datasets formatted in the orthoimage tile system. To ensure geographic diversity and varied urban composition, we use three US areas: Seattle, WA~\cite{seattle_dataset}, Washington D.C.~\cite{dc_dataset}, and Spring Hill, TN~\cite{springhill_dataset}. For Seattle and Spring Hill, the tiles are natively downloaded from the source as 256×256 images, and we adopt zoom level 20 ($\sim$15 cm/pixel), balancing a high enough resolution to visually resolve individual parking spaces and the availability of such data. For DC, the tiles are 512×512 natively, so for objects in the image to achieve the same scale as the other two cities, we use zoom level 19. Then, to unify the image sizes across all downloaded tiles, we resize the DC tiles from 512×512 to 256×256 using the \textit{Python Pillow} library's resize function~\cite{python_PIL} with the \textit{Lanczos} algorithm~\cite{lanczos_algo}. For the bounding coordinates of our downloaded regions, see \autoref{tab:dataset-composition}.

\begin{table}
\begin{tabular}{llrr}
\rowcolor[HTML]{FFFFFF} 
\textbf{Region} & \textbf{\begin{tabular}[c]{@{}l@{}}Bounding\\ Coordinates\end{tabular}} & \multicolumn{1}{l}{\begin{tabular}[c]{@{}l@{}}\textbf{Source}\\ \textbf{Resolution}\end{tabular}} & \multicolumn{1}{l}{\begin{tabular}[c]{@{}l@{}}\textbf{\# images} \\ \textbf{in dataset}\end{tabular}} \\ \hline
\rowcolor[HTML]{FFFFFF} 
Seattle         &  \begin{tabular}[c]{@{}l@{}}(47.9572, -122.4489) \\ (47.4091, -122.1551)\end{tabular}                                                         & 7.62 cm/pixel                                                           & 2,790                                                                     \\
\rowcolor[HTML]{F3F3F3} 
Wash. D.C. & \begin{tabular}[c]{@{}l@{}}(38.9979, -77.1179)\\ (38.7962, -76.9008)\end{tabular}                                                          & 7.62 cm/pixel                                                           & 1,801                                                                     \\
\rowcolor[HTML]{FFFFFF} 
Spring Hill     & \begin{tabular}[c]{@{}l@{}}(35.7943, -87.0034)\\ (35.6489, -86.8447)\end{tabular}                                                          & Unknown                                                                & 534                                                                       \\ \hline
\rowcolor[HTML]{FFFFFF} 
\textbf{Total}  &                                                                                  & \multicolumn{1}{l}{\cellcolor[HTML]{FFFFFF}}                           & \textbf{5,125}                                                                    
\end{tabular}
\caption{Each region represented in our dataset and (1) their bounding coordinates (top is top left coordinate, bottom is bottom right coordinate), (2) their source resolutions, and (3) the number of 512x512 images (tiles).}
\label{tab:dataset-composition}
\end{table}

\begin{figure*}
    \centering
    \includegraphics[width=1\linewidth]{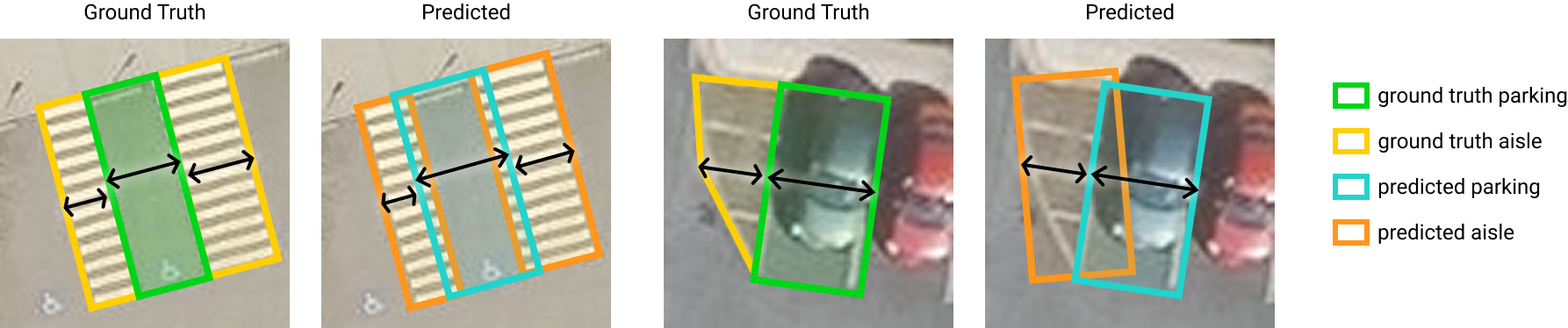}
    \caption{Our definition of width, and a comparison of that definition for ground truth boxes (that can be arbitrary polygons) and predicted boxes (that are restricted to being angled rectangles). Note how the overlap between parking space and access aisle bounding boxes is not double counted.}
    \Description{Definition of width for two parking spaces: one with two rectangular access aisles, and one with one triangular access aisle. Ground truth and predicted boxes are both shown. The prediction for the triangular access aisle is a rectangle that is partially overlapping with the parking space. Arrows indicate the width of each parking space and aisle, in accordance with the definition of width described.}
    \label{fig:defining_width}
\end{figure*}

In Step 2, we stitch together the 256×256 tiles in a 2×2 grid, making 512×512 images, so each image covers a large enough land area to capture several parking spots in one image without being cut off, while balancing resource constraints. The output of Step 2 is over two  million contiguous 512×512 aerial images across our three cities; however, this is too many to manually review and label.

Thus, in Step 3, we created a small seed dataset and trained an initial object detection model (a "hint" model) to guide us towards images that may contain parking. For hint model training, we use a separate open-source aerial image dataset, Denver~\cite{denver_dataset}, to avoid bias towards one of our in-dataset cities. We first manually reviewed Denver's aerial imagery and identified 70 images with parking lots, which we label for disability parking using the same methodology described in section \ref{classes_and_labeling}. These images are 1024×1024, rather than 512x512, as we processed this dataset before defining our broader methodology. We kept this size as they are only used for the hint model, where accuracy requirements are lower.

From this small seed dataset, we train the hint model using YOLOv8~\cite{yolo, ultralytics} for 113 epochs with default settings, which we then run on all two million images from our main cities. Those 512×512 images with a detection confidence higher than 0.3 are marked \sayit{may contain parking}; all others as \sayit{may not contain parking.} We choose 0.3 experimentally: it enables enough flexibility for the hint model to provide false positives, thereby avoiding overfitting. 

Finally, we randomly sample 15,500 images from these two pools (\sayit{may contain parking}, \sayit{may not contain parking}) to again avoid overfitting. From the former, we sample 9,500 images: 6,000 from Seattle, 2,750 DC, and 750 Spring Hill. From the latter, we sample 6,000 images: 4,000 from Seattle, 1,750 DC, and 250 Spring Hill. As the 15,500 images are still composed primarily of null images (images without any disability parking spaces), we further remove a proportion via manual review, so they do not dominate the final dataset. We randomly remove all but $\sim$40\% of the null images from the \sayit{may contain parking} pool, as these are the images that "tricked" the hint model, thereby increasing the robustness of our final dataset. In total, our final dataset includes 5,125 images: 3,065 null and 2,060 with at least one instance of disability parking.

\subsubsection{Classes and Labeling}\label{classes_and_labeling}

Inferring a disability parking spot requires an explicit definition of what constitutes disability parking. Though the ADA clearly defines \textit{what} disability parking is and \textit{how} it should look~\cite{ada_standards_for_accessible_design}, the only required visual indicator is a vertical sign with the international symbol of access. However, such signage is \textit{not} visible from aerial imagery. Although not required, many disability parking spaces also have the access symbol directly painted on the ground, which is discernible using CV methods and aerial imagery. If absent, we must rely on the other visible ADA requirements (\textit{e.g.}, access aisle adjacent to the parking space~\cite{ada_parking_spaces_guidelines}). Thus, in our work, we optimistically consider every space with a visible access aisle as a disability parking candidate, which is also corroborated by our Study 1 participants who emphasized the \textit{utility of an accessible space} whether or not explicitly marked. This biases our dataset towards false positives, which we return to in our Discussion (Section~\ref{discussion_dataset}).

To capture such nuanced definitions of disability parking and to support diverse downstream applications, we define seven classes in our dataset (\autoref{fig:parking_archetypes}): (1) access aisle (\textit{access\_aisle}), (2) curbside (\textit{curbside}), (3) disability parking with no aisles (\textit{dp\_no\_aisle}), (4) disability parking with one aisle (\textit{dp\_one\_aisle}) , (5) disability parking with two aisles (\textit{dp\_two\_aisle}), (6) spaces with one aisle (\textit{one\_aisle}), and (7) spaces with two aisles (\textit{two\_aisle}). \textit{Access aisle} refers to visible non-parking zones adjacent to a parking space. \textit{Curbside parking} denotes spaces that are along the curb of a street and are visibly marked as disability parking. Three classes (\textit{dp\_no\_aisle, dp\_one\_aisle, dp\_two\_aisle}) are spaces that are visibly distinguishable as disability parking, generally via a painted logo, with zero, one, and two access aisles, respectively. The remaining classes (\textit{one\_aisle, two\_aisle}) are spaces without any obvious indication of being disability parking but with adjacent access aisles.

We annotate each object in each image with polygons. Although parking spaces are \textit{generally} consistently sized quadrilaterals, access aisles can be more variable in shape and size, such as triangles, semicircles, or arbitrary polygons. Therefore, to be considered an access aisle (and not simply a striped zone adjacent to a parking space), the aisle must be sufficiently wide and extend along the side of the parking space to a degree where it could aid access in and out of a vehicle. While this typically means along at least $\sim$30\% of the parking space edge, we deferred to the labelers' visual judgment. We include non-labeled examples in \autoref{fig:not-labeled}. Additionally, definitional gaps remain in instances of distortion or atypical parking implementation. In these cases, we again deferred to the labelers' judgment (see \autoref{fig:weird-objects} for examples).

Two members of the research team manually labeled all selected images. Labelers met before, during, and after the labeling process to define and clarify the labeling procedure. Three quarters of the dataset were labeled by researcher A, with the remaining quarter by researcher B. Researcher A then quality checked and adjusted, if necessary, researcher B's labels to ensure consistency. We did not perform an inter-rater reliability test given that all labels were checked for consistency by researcher A.

\begin{figure*}[t]
    \centering
    \includegraphics[width=1\linewidth]{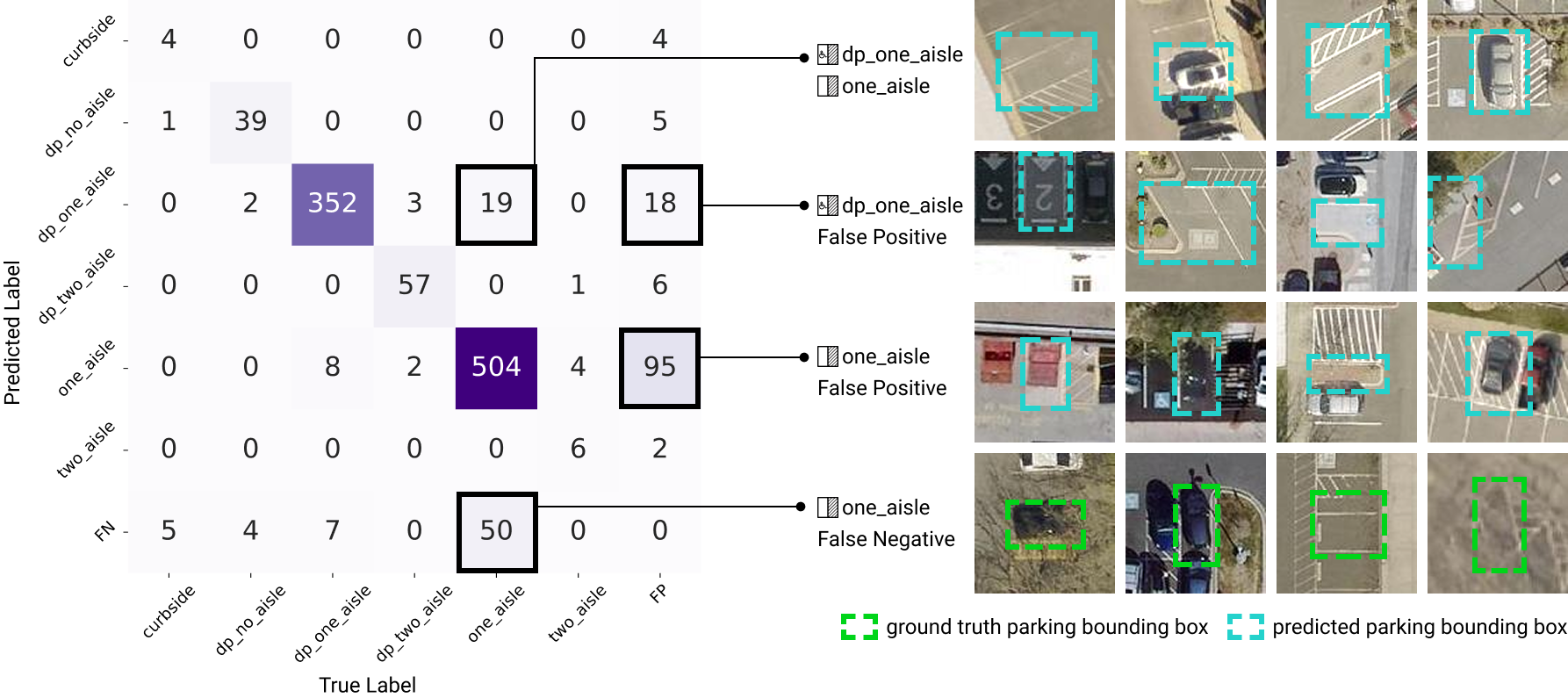}
    \caption{Results from the object detection model on the test set, shown as a confusion matrix, and common examples of the four largest misclassification categories. Predicted bounding boxes are highlighted in cyan, and ground truth bounding boxes (for false negatives) in green. We note certain trends in the misclassifications, including occlusion, smudges, or atypical spaces.
    }
    \Description{A confusion matrix, showing detection accuracy. Most classifications are correct, but four misclassification categories are highlighted: one aisle predicted as dp one aisle, dp one aisle false positives, one aisle false positives, and one aisle false negatives. Example images are shown for each. False positives seem to be spaces that have objects on top of them, or spaces with marks near the bottom where the disability parking logo would be. False negative seems to be occluded or blurred parking spaces.}
    \label{fig:locator-results-confusion}
\end{figure*}

\subsubsection{Summary}

Our final dataset comprises 5,125 images, with 2,060 containing a total of 11,762 labeled objects across all classes (see \autoref{tab:dataset-class-composition} for a breakdown per class). The dataset exhibits significant class imbalance, with one-aisle classes dominating all others, which we discuss further in Section~\ref{discussion}.

\begin{table}[]
\begin{tabular}{lrr}
\rowcolor[HTML]{FFFFFF}
\textbf{Class} & \multicolumn{1}{l}{\cellcolor[HTML]{FFFFFF}\textbf{Quantity in dataset}} & \multicolumn{1}{l}{\cellcolor[HTML]{FFFFFF}\textbf{Percentage}} \\ \hline
\rowcolor[HTML]{FFFFFF}
\textit{access\_aisle} & 4,693 & 39.90\% \\
\rowcolor[HTML]{F3F3F3}
\textit{curbside} & 36 & 0.31\% \\
\rowcolor[HTML]{FFFFFF}
\textit{dp\_no\_aisle} & 300 & 2.55\% \\
\rowcolor[HTML]{F3F3F3}
\textit{dp\_one\_aisle} & 2,790 & 23.72\% \\
\rowcolor[HTML]{FFFFFF}
\textit{dp\_two\_aisle} & 402 & 3.42\% \\
\rowcolor[HTML]{F3F3F3}
\textit{one\_aisle} & 3,424 & 29.11\% \\
\rowcolor[HTML]{FFFFFF}
\textit{two\_aisle} & 117 & 0.99\% \\ \hline
\rowcolor[HTML]{FFFFFF}
\textbf{Total} & 11,762 & 100.00\%
\end{tabular}
\caption{The class composition of our dataset.}
\label{tab:dataset-class-composition}
\end{table}

\subsection{Locating Disability Parking}\label{locating_disability_parking}

We first describe our approach to \textit{locate} disability parking spots followed by \textit{characterizing} its physical dimensions. To detect and locate disability parking spaces in aerial imagery, we employ \textit{multi-class object detection}. Given a 512×512 orthorectified aerial tile, the goal is to predict bounding boxes corresponding to one of six parking-related classes, excluding \textit{access\_aisle}, which we remove during preprocessing because, by definition, this class is associated with at least one adjacent parking space, and its presence is implicitly captured in the class definitions for adjacent labeled spaces, thus not requiring independent prediction.

To examine performance across models, we first benchmarked three candidate object detection models---\textit{YOLOv11 large}~\cite{ultralytics}, \textit{DINO} with a \textit{ResNet50} backbone~\cite{dino}, and \textit{CoDETR} with a \textit{SWIN-L }backbone~\cite{codetr}---the last representing a state-of-the-art transformer-based detection architecture. All models were finetuned on our dataset using a 70/15/15 train/valid/test split. Training was performed on a high performance computing cluster (16 CPU cores, 128 GB RAM, NVIDIA L40S GPU with 48 GB VRAM). Batch sizes were chosen based on model size and hardware constraints: YOLOv11 was trained with a batch size of 16, while the other models used a batch size of 2. Fine-tuning was initialized from publicly released pre-trained weights from each model's official repository (also available in our Hugging Face repository\footnote{\huggingfacepage}). All other hyperparameters, including data augmentation settings, followed default configurations in their respective repositories. Full details, including training time, learning rates, and augmentation strategies, are available in the appendix (\autoref{locator_model_comparisons}) and our GitHub repository\footnote{\githubrepo}. For each model, we selected the best performing checkpoint based on validation set performance, prioritizing high recall to minimize false negatives. This decision reflects the application context, \textit{i.e.}, false positives are easier to manually validate whereas false negatives result in lost detections that are more costly to recover \textit{post hoc}. 

We then evaluate each candidate model on the test set. CoDETR demonstrated the best performance, achieving the highest micro-F1-score (0.89) and recall (0.94)---see \autoref{locator_model_comparisons} for detailed metrics for each model. We thus selected CoDETR for all subsequent analyses. We reuse the results of this experiment for the in-depth evaluation of our locator model's performance (see \autoref{detection_model}), as the application context is the same.

\begin{figure*}
    \centering
    \includegraphics[width=1\linewidth]{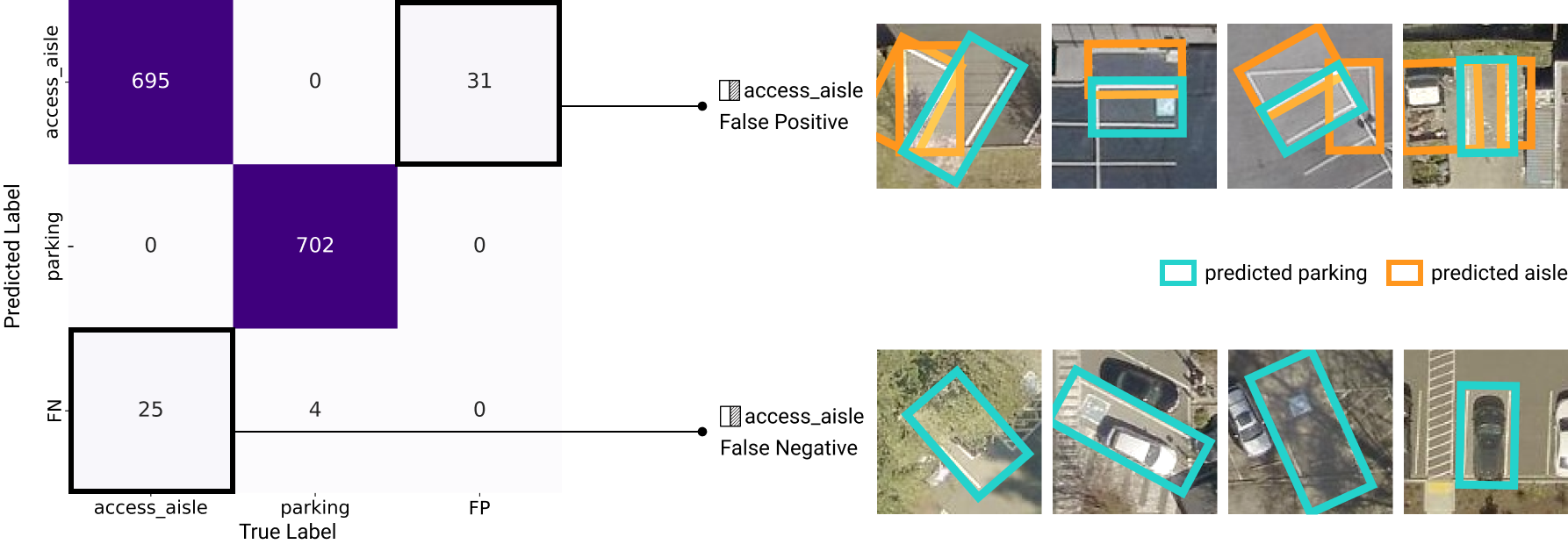}
    \caption{Results from the OBB model on the test set, shown as a confusion matrix, and examples of access aisle false negatives and positives. Predicted OBBs are highlighted in cyan for parking spaces and in orange for access aisles. Ground truth polygons are not shown for false negatives. Once again, we note misclassification trends for occluded and atypical parking spaces.}
    \Description{A confusion matrix showing OBB results in 100x100 cropped images. The model perfectly distinguishes parking spaces from aisles, but there are a number of access aisle false positives and negatives. Examples are shown for each. For false positives, there are duplicate detections or detections of things that are not access aisles. For false negatives, there are occluded spaces by trees or shadows.}
    \label{fig:obb-results-confusion}
\end{figure*}

\begin{table*}[]
\centering
\begin{tabular}{l|l|ll|ll||ll|ll}
\rowcolor{white}
& & \multicolumn{4}{c||}{\textbf{Researcher A vs. Researcher B}} & \multicolumn{4}{c}{\textbf{Researcher A vs. Model}} \\
\rowcolor{white}
& \textbf{Cnt} & \multicolumn{2}{c|}{\textbf{Mean Difference}} & \multicolumn{2}{c||}{\textbf{SD}} & \multicolumn{2}{c|}{\textbf{Mean Difference}} & \multicolumn{2}{c}{\textbf{SD}} \\
\rowcolor{white}
& & \textbf{Pixels} & \textbf{\%} & \textbf{Pixels} & \textbf{\%} & \textbf{Pixels} & \textbf{\%} & \textbf{Pixels} & \textbf{\%} \\
\hline
\textit{null} & 
10 & 
n/a & 
n/a & 
n/a & 
n/a & 
n/a & 
n/a & 
n/a & 
n/a 
\\

\textit{curbside} & 
4 & 
\cellcolor{blues2}-1.03 & 
\cellcolor{blues3}-3.92 & 
\cellcolor{blues2}1.33 & 
\cellcolor{blues4}5.75 & 
\cellcolor{blues3}3.44 & 
\cellcolor{blues6}15.75 & 
\cellcolor{blues3}4.25 & 
\cellcolor{blues7}20.11 \\

\textit{dp\_no\_aisle} & 
12 & 
\cellcolor{blues2}1.94 & 
\cellcolor{blues5}7.37 & 
\cellcolor{blues3}4.84 & 
\cellcolor{blues7}\textcolor{white}{18.22} & 
\cellcolor{blues2}2.47 & 
\cellcolor{blues5}9.11 & 
\cellcolor{blues5}8.63 & 
\cellcolor{blues8}\textcolor{white}{30.74} \\

\textit{dp\_one\_aisle} & 
174 & 
\cellcolor{blues2}1.03 & 
\cellcolor{blues2}2.28 & 
\cellcolor{blues3}4.13 & 
\cellcolor{blues5}9.57 & 
\cellcolor{blues1}0.81 & 
\cellcolor{blues2}2.30 & 
\cellcolor{blues4}5.98 & 
\cellcolor{blues6}13.01 \\

\textit{dp\_two\_aisle} & 
24 & 
\cellcolor{blues1}-0.03 & 
\cellcolor{blues1}0.73 & 
\cellcolor{blues5}7.26 & 
\cellcolor{blues5}9.43 & 
\cellcolor{blues1}0.5 & 
\cellcolor{blues2}1.05 & 
\cellcolor{blues2}2.67 & 
\cellcolor{blues3}4.24 \\

\textit{one\_aisle} & 
139 & 
\cellcolor{blues1}0.16 & 
\cellcolor{blues1}0.91 & 
\cellcolor{blues4}6.80 & 
\cellcolor{blues6}14.47 & 
\cellcolor{blues2}2.98 & 
\cellcolor{blues5}7.65 & 
\cellcolor{blues5}7.62 & 
\cellcolor{blues7}\textcolor{white}{19.36} \\

\textit{two\_aisle} & 
3 & 
\cellcolor{blues2}1.85 & 
\cellcolor{blues2}2.01 & 
\cellcolor{blues3}4.40 & 
\cellcolor{blues4}6.21 & 
\cellcolor{blues2}1.18 & 
\cellcolor{blues2}1.80 & 
\cellcolor{blues2}1.72 & 
\cellcolor{blues2}2.56 \\

\hline
\textbf{Total} & 
366 & 
\cellcolor{blues1}0.65 & 
\cellcolor{blues2}1.77 & 
\cellcolor{blues4}5.57 & 
\cellcolor{blues6}12.05 & 
\cellcolor{blues2}2.01 & 
\cellcolor{blues4}5.40 & 
\cellcolor{blues4}6.91 & 
\cellcolor{blues7}\textcolor{white}{17.52} \\
\end{tabular}
\caption{Results of the error in estimation of width of parking spaces. We compare the difference between two human annotators (Researcher A vs. Researcher B), and the difference between a human annotator and the model (Researcher A vs. Model). Positive indicates overestimation (compared to Researcher A), and negative underestimation. The model sees both larger mean error, and standard deviation of error. Colors indicate magnitude of values using ColorBrewer sequential blue palette.}
\label{tab:width-comparison-obb}
\end{table*}

\subsection{Characterizing Disability Parking}\label{characterizing_disability_parking}

In addition to identifying and locating a disability parking space, we also aim to infer key characteristics such as size. Drawing on our Study 1 results, we found that access aisles and the total number of accessible spaces were key priorities for PwDs. For geometric characterization, we fit \textit{oriented bounding boxes} (OBBs) to each inferred object instance. The OBBs are aligned with the dominant axis of the space, enabling accurate estimations of physical dimensions---particularly width---regardless of the space's orientation in the image. Given their rectilinear structure, parking spaces do not require the granularity of segmentation masks, making OBBs a more efficient and appropriate choice.

To train the OBB model, we again make use of our labeled dataset, however with some preprocessing to align the input data with the OBB model's task. Here, we generate 100×100 crops centered around the centroid of each parking object. In each crop, we extract labels for all objects in the image, adjusting coordinate positions and shapes accordingly (\textit{i.e.}, aligning the coordinates with the new, 100×100 image positions from the old 512×512 image positions). We maintain the same train/valid/test splits as above: any object used for training the locator model was used for training the OBB model. For the test set, we remove any objects close to the image's edge (\textit{i.e.,} that would require padding to reach 100x100), which ensures that the model is tested only on objects where the parking space is whole and not cut-off (for rationale and validity justification, see \autoref{entire_cv_pipeline} and \autoref{overlaptiles}). Our final object-wise train/valid/test splits are 4983/1015/706.

For the OBB model, we use \textit{YOLOv11x}~\cite{ultralytics}. We did not trial other infrastructures, unlike for the locator model, due to YOLOv11x showing positive results at the outset (see \autoref{obb_model_evaluation}). We train with the same compute as above, for 200 epochs, with a batch size of 16, a learning rate of 0.01 with 0.937 momentum and a weight decay of $5\times10^{-4}$. We use all default settings for augmentation, including hue scaling, translation, and horizontal flipping. Details are in the \textit{Ultralytics} documentation.\footnote{\href{https://docs.ultralytics.com/usage/cfg/\#augmentation-settings}{https://docs.ultralytics.com/usage/cfg/\#augmentation-settings}} Further, all specific training values are in our GitHub repository.

\subsection{\sysname{}: An End-to-end CV Pipeline}\label{entire_cv_pipeline}

Finally, we compose the preceding two models---the locator and the characterizer---into an end-to-end pipeline called \textit{\sysname{}} (\autoref{fig:teaser}). Given input bounding coordinates for a region, \sysname{} locates, classifies, and calculates the width of disability parking spaces within that region. The two models function in the same manner as described in Sections \ref{locating_disability_parking} and \ref{characterizing_disability_parking}, with the locator model passing 100×100 crops around the centroid of each detected parking object to the OBB model. Both models use a confidence threshold of 0.3, which were determined experimentally.

To form the final pipeline output, we apply additional logic to the resulting characterizer inferences. Firstly, the OBB model detects all instances of parking spaces and access aisles in an image, however, we are interested in only the center space and its neighboring aisles (as the other spaces have their own separate, independent crops). Thus, we select only the parking space with the highest confidence score that contains the center point of the 100×100 crop. Then, we select aisles that are along at least 40\% of the longest edges of the parking space (\textit{i.e.}, the sides), defined by being within 20 pixels of each other. Second, we derive the width from the resultant OBB detections of the space and its aisles. To calculate the width of a rectangular parking space from its OBB is trivial; however, as an access aisle's OBB's axis can be at an angle relative to the parking space's, what the accessible width of the aisle is is less clear. Therefore, we define the access aisle width as the length of a vector, starting at the midpoint and perpendicular to the parking space edge, pointing away from the centroid, to the farthest intersection point to an access aisle. If there is no intersection point, the width is considered zero. Finally, we calculate the total accessible width for a space by summing the the width of the parking space and the width of any aisles. Since the width of the aisle is defined relative to the parking space, any overlap is \textit{not} double counted (see \autoref{fig:defining_width}).

To format the output, each detected object is georeferenced to a position on Earth, and the results are output to both a JSON file and a shapefile for GIS mapping software.

\subsection{Evaluation and Results} \label{cv_evaluation_and_results}
We individually evaluate both parts of the pipeline, the \textit{locator} and \textit{characterizer}, examining their effectiveness on out-of-sample regions and diverse resolutions, and compare the pipeline's parking width error to the discrepancy between human annotations. 

\begin{table*}[]
\begin{tabular}{llrrrrrr}
\rowcolor[HTML]{FFFFFF} 
\textbf{Region}                       & \textbf{Archetype}     & \textbf{\begin{tabular}[c]{@{}r@{}}True \\ Positives\end{tabular}} & \textbf{\begin{tabular}[c]{@{}r@{}}Ground \\ Truth\end{tabular}} & \textbf{\begin{tabular}[c]{@{}r@{}}Recall \\ (\%)\end{tabular}} & \multicolumn{1}{l}{\cellcolor[HTML]{FFFFFF}\textbf{FN}} & \multicolumn{1}{l}{\cellcolor[HTML]{FFFFFF}\textbf{FP}} & \textbf{\begin{tabular}[c]{@{}l@{}}Mean Width \\ Error (m)\end{tabular}} \\ \hline
\rowcolor[HTML]{FFFFFF} 
Roosevelt (Seattle)                   & Residential            & 43 & 49 & \cellcolor{blues4}87.8 & 6 & 4 & \cellcolor[HTML]{FFFFFF} \\
\rowcolor[HTML]{F3F3F3} 
Northgate (Seattle)                   & Transit Center         & 164 & 174 & \cellcolor{blues6}\textcolor{white}{94.3} & 10 & 39 & \cellcolor[HTML]{FFFFFF} \\
\rowcolor[HTML]{FFFFFF} 
USPS Main Office (DC)                 & Urban Commercial       & 117 & 122 & \cellcolor{blues7}\textcolor{white}{95.9} & 5 & 11 & \cellcolor[HTML]{FFFFFF} \\
\rowcolor[HTML]{F3F3F3} 
Audi Field (DC)                       & Sports Stadium         & 41 & 43 & \cellcolor{blues7}\textcolor{white}{95.3} & 2 & 8 & \multirow{-4}{*}{\cellcolor[HTML]{FFFFFF}-0.20 (SD=1.63)} \\ \hline
\rowcolor[HTML]{FFFFFF} 
Koreatown (Los Angeles)               & Commercial/Residential & 79 & 123 & \cellcolor{blues1}64.2 & 44 & 23 & \cellcolor[HTML]{F3F3F3} \\
\rowcolor[HTML]{F3F3F3} 
Del Amo Fashion Center (Torrance, LA) & Urban Commercial       & 163 & 227 & \cellcolor{blues2}71.8 & 64 & 30 & \multirow{-2}{*}{\cellcolor[HTML]{F3F3F3}-0.29 (SD=1.63)} \\ \hline
\rowcolor[HTML]{FFFFFF} 
Shopping Mall (Waltham, MA)           & Suburban Commercial    & 132 & 164 & \cellcolor{blues3}80.5 & 32 & 20 & 0.53 (SD=1.16) \\ \hline
\rowcolor[HTML]{E6E6E6} 
\textbf{Total}                        & \textbf{All Regions}   & \textbf{739} & \textbf{902} & \cellcolor{blues3}\textbf{81.9} & \textbf{163} & \textbf{135} & \textbf{-0.10 (SD=1.58)} 
\end{tabular}
\caption{The seven one km$^{2}$ regions we evaluated our pipeline on. We do not distinguish between categories for parking spaces detected, only if they were detected at all. Mean Width Error is shown, \textit{in meters}, aggregated across three categories: regions represented in our training data (Seattle, D.C.), Los Angeles, and Massachusetts. Again, positive number indicates overestimation, and negative underestimation. We see that the model consistently detects a higher proportion of objects, and with a lower error for width, for the regions represented in the training set.}
\label{tab:region-evaluation}
\end{table*}

\subsubsection{Detection Model}\label{detection_model} To examine overall performance, we evaluated the CoDETR disability parking locator model on the test set (717 images across Seattle, DC, and Spring Hill). We matched detected objects to ground truth using the \textit{Hungarian} method for bipartite matching ~\cite{hungarian_method}, with an \textit{Intersection Over Union} (IOU) threshold exceeding 0.5 to ensure optimal correspondence. For all aggregated performance statistics (precision, recall, and F1 score), we use the micro-average~\cite{sokolova_f1score} to capture the holistic performance of the model across classes. However, given the dataset imbalance, we encourage readers to observe the per-class statistics for a comprehensive view. Overall, the CoDETR disability parking locator model performs well with an overall precision of 0.85, recall of 0.94, and F1 score of 0.89---see class-specific statistics in \autoref{tab:locator-results}. 

\textbf{Qualitatively examining errors.} To better understand model performance, we manually examined all misclassifications, false negatives (where a parking space that exists is not detected), and false positives (where null pixels are detected as a parking space)---\autoref{fig:locator-results-confusion}. For misclassifications (\textit{N=31}), the most common error is \textit{one\_aisle} misclassified as \textit{dp\_one\_aisle} or vice versa, perhaps since the only difference from the aerial view is a visual indicator (\textit{e.g.}, painted logo). In fact, common cases of \textit{dp\_one\_aisle} being predicted as \textit{one\_aisle} are when the logo is faded, smudged, or partially occluded. Inversely, in many cases, a space that is occluded or has some mark near the bottom, is misclassified as \textit{dp\_one\_aisle}.

\begin{table}[b]
\centering
\renewcommand{\arraystretch}{1.1}
\setlength{\tabcolsep}{3pt}
\begin{tabular}{@{}lrrrr@{}}
\rowcolor[HTML]{FFFFFF}
\textbf{Class} & \multicolumn{1}{l}{\cellcolor[HTML]{FFFFFF}\textbf{\# in test set}} & \multicolumn{1}{l}{\cellcolor[HTML]{FFFFFF}\textbf{Precision}} & \multicolumn{1}{l}{\cellcolor[HTML]{FFFFFF}\textbf{Recall}} & \multicolumn{1}{l}{\cellcolor[HTML]{FFFFFF}\textbf{F1}} \\
\midrule
\textit{curbside} & 10 & \cellheatmapcolor{0.50}0.50 & \cellheatmapcolor{0.40}0.40 & \cellheatmapcolor{0.44}0.44 \\
\textit{dp\_no\_aisle} & 45 & \cellheatmapcolor{0.87}0.87 & \cellheatmapcolor{0.87}0.87 & \cellheatmapcolor{0.87}0.87 \\
\textit{dp\_one\_aisle} & 367 & \cellheatmapcolor{0.89}0.89 & \cellheatmapcolor{0.96}0.96 & \cellheatmapcolor{0.93}0.93 \\
\textit{dp\_two\_aisle} & 62 & \cellheatmapcolor{0.89}0.89 & \cellheatmapcolor{0.92}0.92 & \cellheatmapcolor{0.90}0.90 \\
\textit{one\_aisle} & 553 & \cellheatmapcolor{0.82}0.82 & \cellheatmapcolor{0.88}0.88 & \cellheatmapcolor{0.85}0.85 \\
\textit{two\_aisle} & 10 & \cellheatmapcolor{0.75}0.75 & \cellheatmapcolor{0.55}0.55 & \cellheatmapcolor{0.64}0.64 \\
\midrule
\textbf{Total} & \textbf{1,047} & \cellheatmapcolor{0.85}\textbf{0.85} & \cellheatmapcolor{0.94}\textbf{0.94} & \cellheatmapcolor{0.89}\textbf{0.89} \\
\bottomrule
\end{tabular}
\caption{Results of our locator model, by class. Note that for the Total results, misclassifications, or instances where an object of one class is predicted as another, are considered false positives for the statistics calculation. Performance is high for classes with high representation, but lower for classes that are underrepresented (\textit{curbside}, \textit{two\_aisle}). Total, aggregated statistics are formed from the micro-average of the class level statistics.}
\label{tab:locator-results}
\end{table}

\begin{figure*}[t]
  \includegraphics[width=\textwidth]{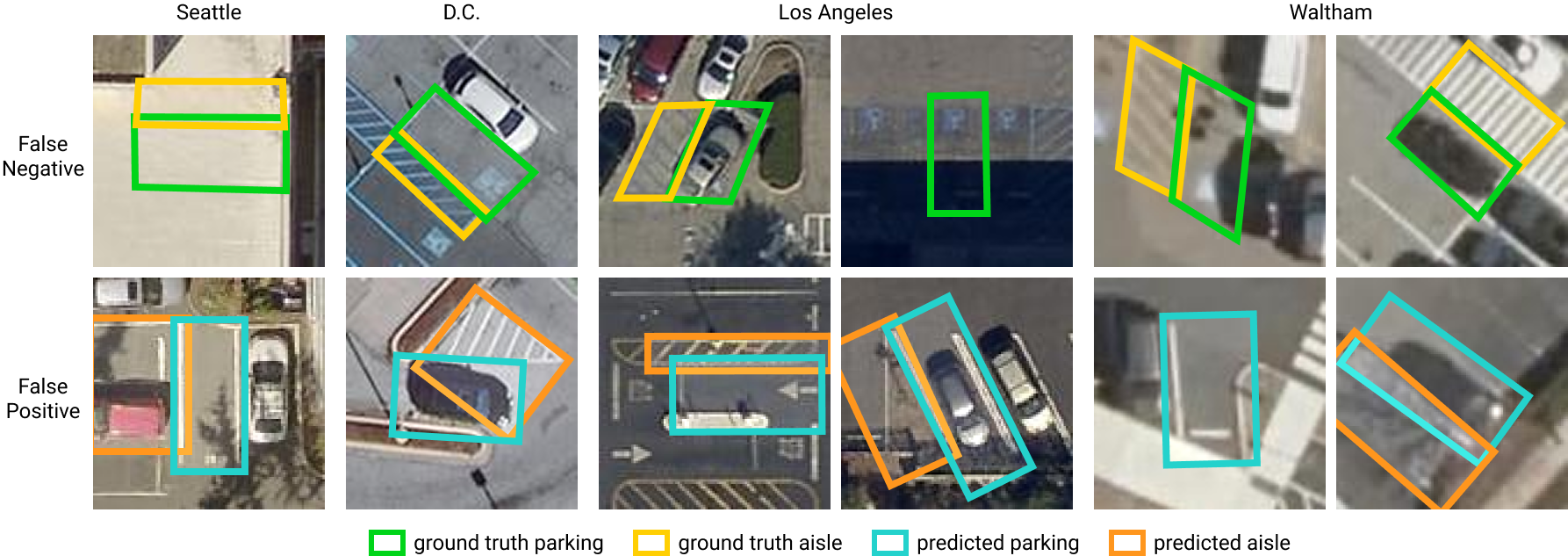}
  \caption{Example FNs and FPs for each evaluation region with ground truth annotations and predictions shown. We see similar error patterns as before; however, LA disability parking demarcations are often blue (rather than white or yellow) and more faded, resulting in lower contrast with the pavement. Waltham data is at a much lower resolution, resulting in blurriness.}
  \Description{A grid of images of parking, showing examples for false negatives and false positives for each evaluation region: Seattle, Washington DC, Los Angeles, and Waltham, Massachusetts. False negatives are low contrast or low definition spaces. False positives are spaces with objects next to them that may look like striped zones, but are not, or, are roads with a striped zone adjacent.}
  \label{fig:region-evaluation-examples}
\end{figure*}

In examining \textit{false negatives} (\textit{N=66}), common patterns are paint fade (thus lower contrast with the pavement), occlusion (\textit{e.g.}, from trees or shadows), or an access aisle not being detected (thus, not identifying the space as a \textit{one\_aisle}). For \textit{false positives} (\textit{N=130}), common patterns are access aisles that were blocked by an object, spaces or aisles that were not labeled (due to occlusion or distortion), or spaces that looked similar to a parking space with a logo (\textit{e.g.,} an empty area with an adjacent striped zone, or a number at the bottom). Interestingly, in the former two cases, perhaps a more lenient labeling scheme may have identified the spaces, making the false positives correct detections.

Finally, single parking spaces detected multiple times with different classifications (with one being correct) contributes to the error rate. This is avoided for the full pipeline by filtering overlapping detections by confidence score.

\subsubsection{OBB Model}\label{obb_model_evaluation}

Beyond detections, we also examine the other key component of our pipeline, the \textit{characterizer} model. Here, we evaluate it as we did the detection model: on the (100×100) test set, matching objects with the Hungarian method and an IOU of 0.5. However, we compute only whether the space was detected as an access aisle or parking space and the presence of false negatives/positives, because the OBB model is solely used for width estimation, with class prediction handled by the detection model.

Overall, similar to our detection model, we see high performance, with only 29 false negatives and 31 false positives, of 1,426 detected objects---see \autoref{fig:obb-results-confusion}. Our space characterizer perfectly distinguishes access aisles from parking spaces. However, access aisles experience some false negatives (3.5\% not detected), often due to occluded, faded, or smudged aisles, and false positives (4.3\% of total access aisle detections) that commonly result from duplicate detections of non-rectangular aisles, or the misidentification of neighboring spaces or striped zones.

\textbf{Evaluating width characterization.} In detecting parking spots and their aisles in a cropped image as an OBB, the key responsibility of the characterizer is to quantify parking space width (including access aisles), via the dimensions of the OBBs. To evaluate width characterization, we compared our model's error to the error between two human researchers' labels. Researchers A and B labeled the same, randomly sampled 356 object subset of the test set (plus 10 randomly sampled false positives from the detection model evaluation, as null trials), and we calculated the differences in width from their respective labels. As location data (which is required to derive real-world distance) was lost during the test set preprocessing, all measurements for evaluation are in pixels, not real-world units. 

Overall, we find that the model performs well: only slightly worse than the humans with an average error of 5.40\% \textit{vs.} 1.77\% and overall standard deviation of 17.52\% \textit{vs.} 12.05\%, though performance depends on class---see~\autoref{tab:width-comparison-obb} and \autoref{width_characterizaion_histograms}.

\subsubsection{Cross City Performance}\label{cross_city_performance}

Finally, to investigate how well \sysname{} generalizes to regions outside of those used to train the models, we selected seven additional one km$^2$ regions, four from our original cities that were not in our training set (two from Seattle; two from DC) as well as two neighborhoods in Los Angeles~\cite{la_dataset} and one from Massachusetts~\cite{mass_dataset}. These selections were based on geographic and zone type variation in addition to data availability in Tile2Net. Aerial imagery from Massachusetts has a much lower resolution (15 cm/px \textit{vs.} 7.6 cm/px), which also enabled us to examine performance on lower resolutions. For ground truth, one member of the research team manually labeled all parking spaces and aisles with the same seven classes as before, and we release these additional seven regions and labels to our HuggingFace. We evaluate the entire pipeline: the ability to detect parking spaces, and the accuracy of the pipeline's estimation of their widths. Each detected parking space was matched with a ground truth label for evaluation, again with the Hungarian bipartite matching method based on an IOU of 0.3. We report results for \textit{total objects detected}, without distinguishing between parking classes, due to the low cross-class sample count.

Overall, as expected, detector model performance drops but still performs fairly well, particularly on the neighborhoods drawn from the same cities as the training set (Seattle and DC) with average recall above 90\%; however, performance in LA and Waltham, MA drops significantly: to $\sim$69\% in LA and 81\% in MA, which suggests some amount of per-city training data may be necessary. To advance understanding of model performance, we once again qualitatively analyzed errors in each region (\autoref{fig:region-evaluation-examples}); we find similar error patterns as reported in Sections \ref{detection_model} and \ref{obb_model_evaluation}. However, we note two notable distinctions with LA parking data: a prevalence of blue paint for disability parking  (\textit{vs.} yellow or white), resulting in reduced contrast with the surrounding pavement, and more frequent and severe paint fading, possibly attributable to intense sunlight. The human labeler also found LA data more difficult to label than other regions given these reasons, which may have contributed to the overall error rate (given that the labels themselves may not be 100\% accurate). These patterns may also have contributed to the width error rate, as parking spaces where the borders are not clearly defined are more imprecisely detected. 
\begin{figure*}
    \centering
    \includegraphics[width=1\linewidth]{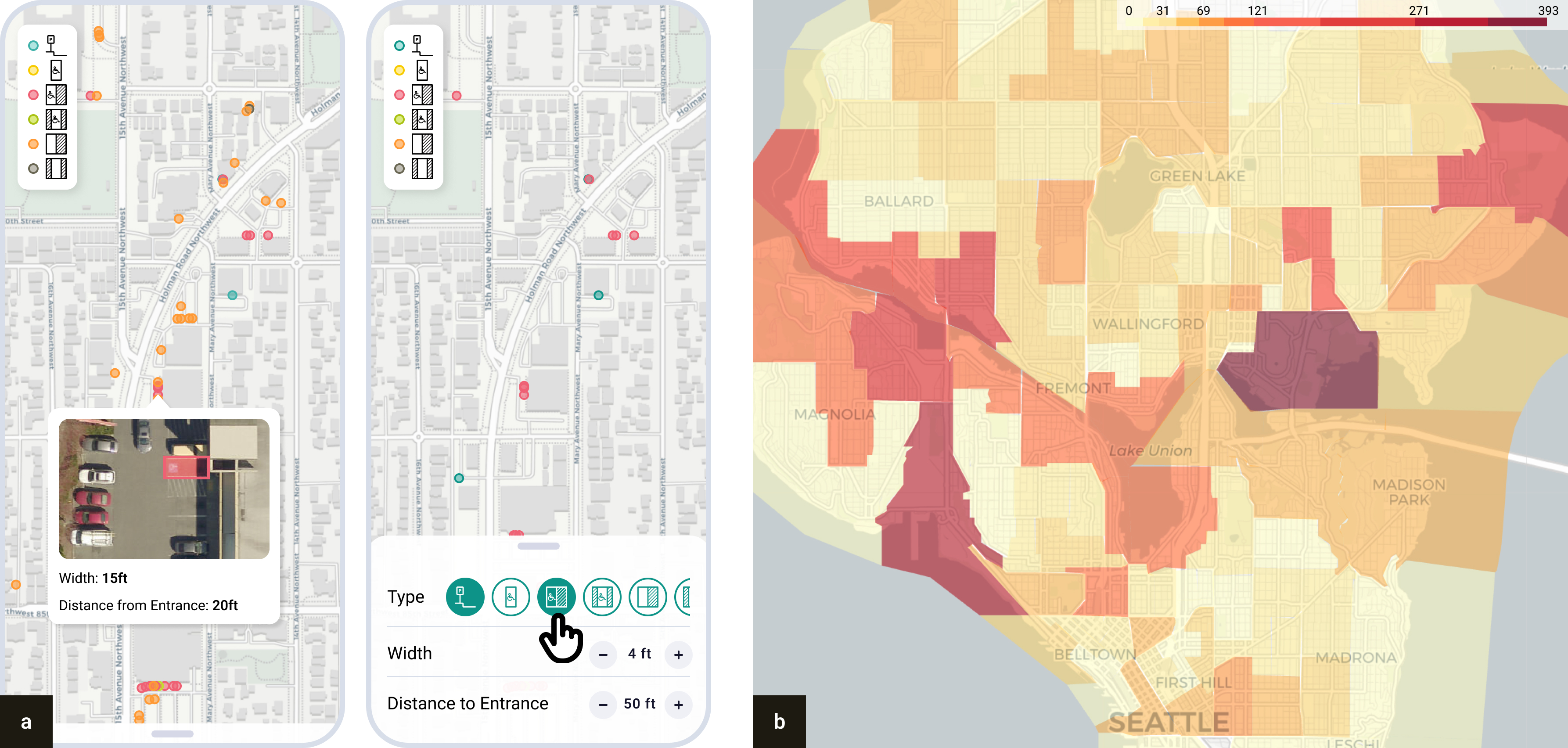}
    \caption{Two applications utilizing the results from our detection tool: (a) a personalized parking search and filter application, and (b) an urban analytics visualization displaying number of disability parking spaces per census tract in Seattle.}
    \Description{On the left, two images of a phone application, showing a map with colored dots indicating different types of parking spaces. A pop-up window shows an aerial view of a selected parking space. A menu provides the option to filter by parking class, width, and distance to entrance. On the right, a choropleth visualization showing number of parking spaces per census tract in Seattle.}
    \label{fig:case-study-apps}
\end{figure*}

\section{Case Study Applications}

To help demonstrate the utility of \sysname{} as well as our findings from Study 1, we present two case study applications: a personalized disability parking app (a lo-fi mockup) and an interactive urban analytics visualization (a mid-fi prototype). 

\subsection{Personalized Parking Search and Filter}

As highlighted in Study 1, individuals with disabilities (PwDs) highly value pre-trip information regarding location accessibility, as it significantly reduces travel time and anxiety. Regarding parking, the availability and size of spaces were frequently cited as crucial factors. Leveraging classifications and width data from \sysname, we designed a lo-fi mapping application (\autoref{fig:case-study-apps}a) that identifies accessible parking options within a given area. The tool allows users to refine their search based on: the number of adjacent aisles, the total width of the parking space, and its proximity to their destination. For example, a user operating a ramp-equipped van, who prioritizes space over immediate proximity to an entrance, can effectively filter for suitable parking spots and plan their travel accordingly. While the application does not provide real-time occupancy data (a limitation of our pipeline), it empowers users to locate appropriately sized parking spaces that meet their specific vehicle and accessibility requirements, thereby facilitating the identification of suitable alternatives if a preferred space is occupied.

\subsection{Disability Parking Analytics}

As stipulated by the Americans with Disabilities Act (ADA), 4-8\% of total public parking must be designated as accessible parking~\cite{laban_dp_privilegeorright}. To help disability advocates, government officials, and even end-users themselves understand the distribution and density of disability parking in their communities, we designed and developed a mid-fi interactive disability parking visual analytics tool. Illustrated in Figure \ref{fig:case-study-apps}b, the tool enables users to conduct in-depth investigations into disability parking trends, which can be further disaggregated by community, zoning regulations, and business concentration. Such visual analytics provide a data-driven foundation for local governments to formulate and revise parking and zoning policies. Furthermore, these visualizations could be integrated into other regional accessibility metrics, such as \textit{AccessScore}~\cite{li_accessibility_2025}, empowering individuals to make informed decisions regarding travel and residential choices based on accessibility considerations.

\section{Discussion} \label{discussion}
In this paper, we explored the nuances of disability parking in the US, sharing formative findings on its perceived benefits, uses, and drawbacks. We also developed a new algorithmic pipeline to detect and assess disability parking spots from aerial imagery, created a novel disability parking dataset, and showcased two practical applications of our methods. Below, we reflect the implications of our work, our dataset contributions, the future of CV applied to disability parking, and key limitations and future work.

\subsection{Implications for Design}

Drawing on our study findings, we synthesize five key recommendations for the design of interactive apps for disability parking.

\begin{itemize}
    \item \textbf{Beyond location.} While showing the location of disability parking spots on a map is a critical accessibility advancement, it is insufficient to support diverse mobility needs, vehicle types, and the various requirements for access aisles. Future apps should leverage \sysname{}'s ability to infer characteristics like the presence and width of aisles and surface related metadata to the user (similar to \autoref{fig:case-study-apps}a).
    \item \textbf{Personalized filtering.} Because preferences for disability parking vary significantly---as a function of ambulation ability, use of mobility aid, vehicle type---apps should empower users to filter for their specific accessibility criteria (\textit{e.g.,} minimum width, presence and orientation of access aisles).
    \item \textbf{Comprehensive routing.} While finding an accessible parking spot is critical so too is the journey from the parking to the building entrance. Future tools should integrate assessments of the "out-of-car" experience, including the presence of sidewalks, pathway accessibility, and distances.
    \item \textbf{Trust and transparency.} Our participants expressed skepticism about the reliability of disability parking information, partly due to issues like illegal occupancy and inconsistent implementation of regulations. To build trust, apps should surface information sources and, if possible, allow users to actually look at raw image sources (for verification). 
    \item \textbf{Real-time occupancy.} The ideal app would also provide real-time availability data, which is beyond the capabilities of our techniques. Future work should explore hybrid solutions connected to real-time sensors.
    \item \textbf{Human-in-the-loop.} While our CV techniques perform well, they are imperfect. Apps should empower users to provide \textit{in situ} insights on the quality of a parking spot and provide feedback on CV detections via their lived experience.
\end{itemize}

\subsection{Dataset Contribution}\label{discussion_dataset}

A key contribution of our work is the first open-source, human-labeled dataset of disability parking and access aisles from aerial orthoimages. This dataset comprises 11,762 labeled objects across 5,125 images from three distinct US regions: Seattle, Washington D.C., and Spring Hill, TN. By stratifying "accessible parking" into seven distinct classes, including those with varying numbers of aisles and different visual indicators (\textit{e.g.,} painted logos), we enable fine-grained analysis and customization for both macro-scale urban planning and micro-scale personalization. Our dataset includes a significant number of null examples (3,065 images) from the same regions, allowing researchers to filter null proportions appropriate for their desired use case. We publish the dataset in the commonly used COCO format~\cite{lin_microsoft_2014} for convenience.

While the dataset exhibits class imbalance, with one-aisle classes dominating, this imbalance likely reflects real-world conditions, as ADA guidelines do not mandate two aisles for all accessible spaces: only "van accessible" spaces are required to have two aisles and curbside disability parking is not mentioned~\cite{ada_parking_spaces_guidelines}. In our locator model, we find that this imbalance does not significantly affect results (beyond the curbside class, which only has 36 samples). However, future dataset expansions should include efforts to reduce the imbalance.

\subsection{Disability Parking CV: A New Benchmark}

Our novel computer vision pipeline, composed of a locator and a characterizer model, sets a new benchmark for scalably detecting and characterizing disability parking from aerial imagery. The pipeline achieves a micro-F1 score of 0.89 for detection, demonstrating high overall performance, particularly for classes with strong representation in the dataset. Importantly, our characterizer model estimates the width of parking spaces with an average error of only 5.4\%, which directly helps assess compliance with accessibility standards and to provide actionable data to PwDs, a need strongly voiced by our interview participants.

The open-sourcing of our pipeline code and experiments alongside our dataset helps advance research in CV-based analyses of accessible urban infrastructure~\cite{hosseini_mappingthewalk, larkin_builtenvironment_gsv_satellite, kulkarni_busstopcv, duan_scalingcrowd_assets22, liu_finegrain_assets24} and provides foundational benchmarks for this emerging area. While our pipeline excels at identifying clearly defined and visible disability parking spaces, it struggles with "atypical" parking spaces---just as our human annotators did---due to factors such as faded paint, blurriness, and unexpected design variation. Our decision to prioritize lowering false negatives over false positives reflects the application context, where undetected accessible parking is more costly than manually verifying a false positive. This approach aligns with the principle that humans, particularly those with lived experience, should remain in the loop for validating and verifying algorithmic outputs related to accessible urban infrastructure.

\subsection{Scale and Generalizability}

Our longterm, overarching goal is to accurately detect, characterize, and track every disability parking spot in the US, if not the world. While our experiments demonstrate high performance, particularly for regions within our training set, there are two key limiting factors: (1) in \autoref{cross_city_performance}, we show that performance drops without some per-city training---future work should more comprehensively examine performance as a function of training set size and diversity; (2) the availability of high-resolution aerial orthoimagery, which are often created by public agencies like the US Geological Survey~\cite{usgs_geo_survey} or local governments (the Seattle~\cite{seattle_dataset}, D.C.~\cite{dc_dataset}, Massachusetts~\cite{mass_dataset}, and LA~\cite{la_dataset} datasets are all provided on open gov city websites). Continued expansion of the dataset across a higher diversity of regions, resolutions, and zoom levels would bolster model generalizability and potentially reduce the need for extensive fine-tuning on deployment regions.

\subsection{Limitations}\label{limitations}

Our work has several primary limitations. First, in Study 1, our participant pool was almost entirely composed of wheelchair users. Greater diversity of disability type and mobility aid would enhance the generalizability of our qualitative findings. Second, aerial imagery inherently limits the scope of detectable parking, as indoor parking facilities (\textit{e.g.,} in parking garages) are not visible. Third and relatedly, not all disability parking can be identified from aerial imagery, including in instances where a space lacks an access aisle and is only visually denoted by a vertical sign. Furthermore, environmental factors like occlusion from trees or shadows can impact accuracy. Fourth, our approach is reliant on the availability and quality of high-resolution orthoimagery. Even in cases where such data \textit{is} available, aberrations within the dataset can affect the results (\textit{e.g.,} in DC, the entire National Mall is blurred, presumably for security reasons). Similarly, real-time applications are also not possible using only orthoimage datasets, given they are typically only updated after several years, if ever. Finally, our definitions of "access aisle" and "width" may not be appropriate in all scenarios. For example, in our dataset we choose to label striped, triangular wedges between two parking spaces as an access aisle; however, if two long, parked cars were to park in the adjacent spaces, there would be no path out from the aisle.

\subsection{Future Work}

Future work should address the limitations outlined above. Employing multiple aerial orthoimage datasets of the same region, captured at different times, could help mitigate issues caused by parked cars occluding disability parking features. Integrating non-aerial imagery sources, such as Google Street View, could enable the identification of spaces marked only by vertical signage and potentially provide visibility into indoor parking areas or spaces entirely occluded from a bird's-eye view. Further expansion of the dataset to include a higher diversity of regions, resolutions, and zoom levels would likely bolster model generalizability and may reduce the need for extensive fine-tuning on new deployment regions. 

Beyond addressing limitations, several potential extensions could improve \sysname{}. First, additional classification models or ensemble voting approaches~\cite{ganaie_ensemblemethods} could be used to reduce the number of false positives. Second, additional characterizer models could be added along the pipeline that accept the same image input shape as the OBB model, such as quantifying a parking space's degree of paint fade or if it is angled. Finally, incorporation of contextual information, such as obstructions on the path to the entrance or distance from traffic, would allow us to make more holistic claims on the accessibility of a given parking space.

We envisioned two, non-exhaustive use cases for the data produced by our pipeline. More applications and tools could also benefit from the disability parking detections. Disability parking data could be combined with sidewalk data~\cite{saha_projectsidewalk, li_accessibility_2025}, which could then be integrated into a personalized routing app, or to generate a community level accessibility score. Further, a tool could be created that audits and notifies individual businesses if they are compliant with ADA  parking guidelines~\cite{ada_parking_spaces_guidelines} based on quantity and width. 

\section{Conclusion}

Through mixed-methods research, including an interview study of 11 PwDs and the development and evaluation of a CV pipeline that can locate and characterize disability parking spaces, we contribute:

\begin{enumerate}
  \item Design and policy guidelines for the implementation and maintenance of disability parking.
  \item An open-source dataset of 11,762 labeled disability parking spaces and access aisles that can be used for classification, detection, segmentation, or other analysis tasks.
  \item A CV tool that locates, classifies, and characterizes the width of disability parking from aerial orthoimagery.
  \item Two envisioned applications that leverage the preceding contributions to enable large-scale disability parking auditing and individual-scale accessibility filtering.
\end{enumerate}

Our work highlights the potential for large-scale, computer vision disability parking detection and characterization tools to improve both the implementation of disability parking at large and for the data from such tools to bolster government-level analysis and individual-level customization. 
Our work showcases both the need for such investigative and personalized tools and the ability for computer vision to serve as a vital step towards that end.

\begin{acks}
We thank all study participants as well as Adela Mu for urban planning and GIS guidance, Daniel Campos Zamora and Charles DeLorey for their local expertise on our test regions, academic writing advisor Sandy Kaplan, UW CREATE, and CREATE community engagement manager Kathleen Quin Voss. This work was supported by NSF grant \#2411222 and the US DOT National Center for Understanding Future Travel Behavior and Demand (\#69A3552344815 and \#69A3552348320). \autoref{fig:teaser} icons by~\cite{wheelchair_icon, bsdstudio_icons}.
\end{acks}

\bibliographystyle{ACM-Reference-Format}

\bibliography{references}

\appendix\appendix
\section{Appendix}

\subsection{Detecting Objects Split Across Tiles}\label{overlaptiles}

\begin{figure}[h]
    \includegraphics[width=\linewidth]{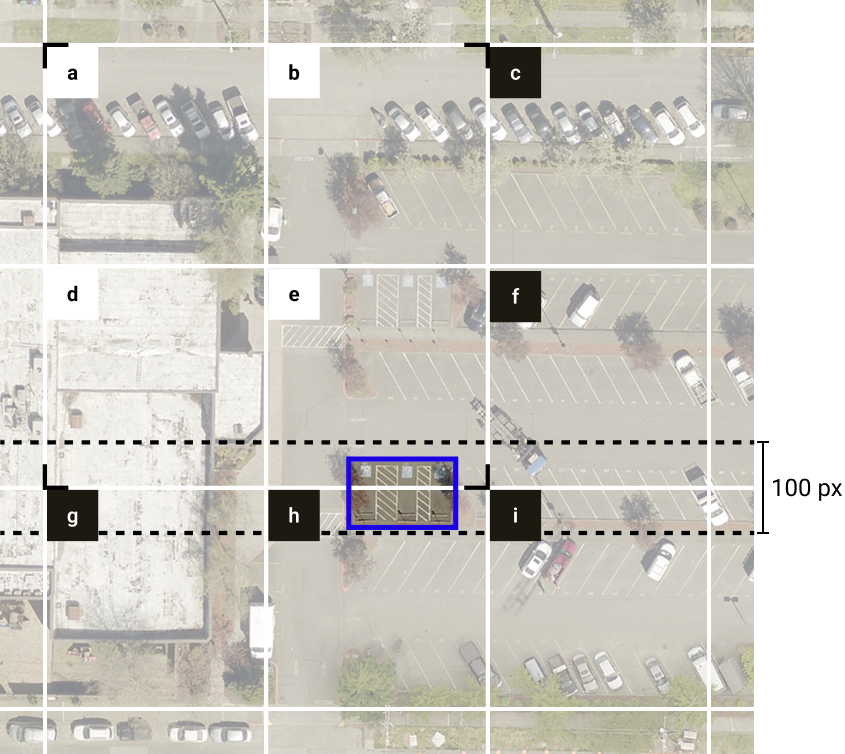}
    \caption{Illustration for Algorithm~\ref{alg:overlap_algo}. We employ a sliding-window algorithm to detect objects split between tiles. In this example, the input to the pipeline is the top left tile position \{\textit{a}\}, from which it will stitch together and detect objects in the 2x2 square \{\textit{a}, \textit{b}, \textit{d}, \textit{e}\}, denoted in white. The parking spaces in blue are half in and half out of the square. As the algorithm proceeds, the parking spaces will be detected in their entirety in the squares \{\textit{d}, \textit{e}, \textit{g}, \textit{h}\} and \{\textit{e}, \textit{f}, \textit{h}, \textit{i}\} and be included in the results for the square \{\textit{a}, \textit{b}, \textit{d}, \textit{e}\}.}
    \Description{A three by three grid of aerial images of a parking lot. The four tiles making a square in the top left are highlighted in white, the rest in black. There are three parking spaces highlighted in blue, between the boundary between the center tile and the tile below it. There are horizontal dotted lines above and below them, and the distance between them is 100 pixels.}
    \label{fig:overlap_algo}
\end{figure}

\begin{algorithm}[h]
\caption{Detect objects in a 512×512 Tile (refer to \autoref{fig:overlap_algo})}\label{alg:overlap_algo}
\begin{algorithmic}
\State \textbf{Time Cost:} $O(4n)$.
\State \textbf{Assume:} Objects are no wider than 100 pixels.
\State \textbf{Input:} Top left tile position \{\textit{a}\}.
\State \textbf{Step 1:} Detect in \{\textit{a}, \textit{b}, \textit{d}, \textit{e}\}. Ignore objects with a centroid <50 px from border.
\State \textbf{Step 2:} Detect in \{\textit{b}, \textit{c}, \textit{e}, \textit{f}\}. Ignore objects with a centroid <50 px from border and >50 px from the vertical middle.
\State \textbf{Step 3:} Detect in \{\textit{d}, \textit{e}, \textit{g}, \textit{h}\}. Ignore objects with a centroid <50 px from border and >50 px from the horizontal middle.
\State \textbf{Step 4:} Detect in \{\textit{e}, \textit{f}, \textit{h}, \textit{i}\}. Only keep objects with a centroid within a 100 px square around the center point.
\State \textbf{Return:} All non-ignored objects and their positions.
\end{algorithmic}
\end{algorithm}

Because of tiling, some objects will be split across tiles and potentially detected multiple times. We consider two possible solutions to this problem: (1) to combine detected objects across borders into one based on their geometry (the approach used by Tile2Net~\cite{hosseini_mappingthewalk}), or (2) to create a new image, encapsulating the area at the border, and to detect within this new image, thereby detecting one whole object rather than multiple fragments of an object. 

We opted for Method 2 as Method 1 would raise other detection problems to consider. For example, parking spaces are quite small relative to the size of the image;  what  if, by being split across tiles, a space is only detected in one tile or in no tiles at all? Furthermore, a parking space's classification significantly depends on its features (\textit{e.g.}, the number of aisles it has, if it has a painted logo, \textit{etc.}). What if the space is split so that the logo or an aisle is visible in one tile but not the other? How would the class be determined? Finally, parking spaces tend to be both small and adjacent to one another. The object combination would need to be very robust to avoid combining two neighboring objects into one.

We avoid all these issues by using Method 2, ensuring we detect only \textit{whole} objects, but at the cost of computation. To do this, we assume that a parking space (with aisles) is no wider than 100 pixels in a given dimension, which, at our zoom level, holds for all but the most extreme of scenarios (such as semi-truck parking). In this way, the locator model is only tasked with detecting entire parking spaces and passes only unpadded, 100×100 crops to the OBB model. For a detailed breakdown of the approach and algorithm, including computation cost, see \autoref{fig:overlap_algo} and Algorithm \ref{alg:overlap_algo}.

\subsection{Georeferencing}\label{georeferencing}

For a pixel in a given tile whose $x$ and $y$ positions are in the orthoimage tile system, we can obtain that pixel's location on Earth (given zoom level 20). Orthoimage tiles are representations of the common Web Mercator projection~\cite{webmercator_projection}. We discern a pixel's latitude and longitude (in the commonly used WGS84~\footnote{\href{https://epsg.io/4326}{https://epsg.io/4326}}) with the following:

$$
longitude = \frac{x}{2^{20}}\cdot360 -180
$$
$$
latitude = \frac{180}{\pi}\cdot\arctan(\sinh(\pi -2\pi\frac{y}{2^{20}}))
$$
Here, $x$ and $y$ are the fractional values of the pixel in relation to the tile in which it belongs. For example, in a 256x256 orthoimage tile with $x=168046, y=366004$, if a pixel is at the 128th row and column, the values used in the preceding equations are $x=168046.5,y=366004.5$. Then, to calculate distances, and therefore widths, the latitude and longitude are converted into points in a projected coordinate system, EPSG:3857\footnote{\href{https://epsg.io/3857}{https://epsg.io/3857}} (\textit{i.e.}, pseudo-mercator with units in meters), and simple Euclidean distance is used.

\newpage

\subsection{Width Characterization Histograms}\label{width_characterizaion_histograms}

\begin{figure}[ht]
    \centering
    \onecolumn
    \includegraphics[width=\textwidth]{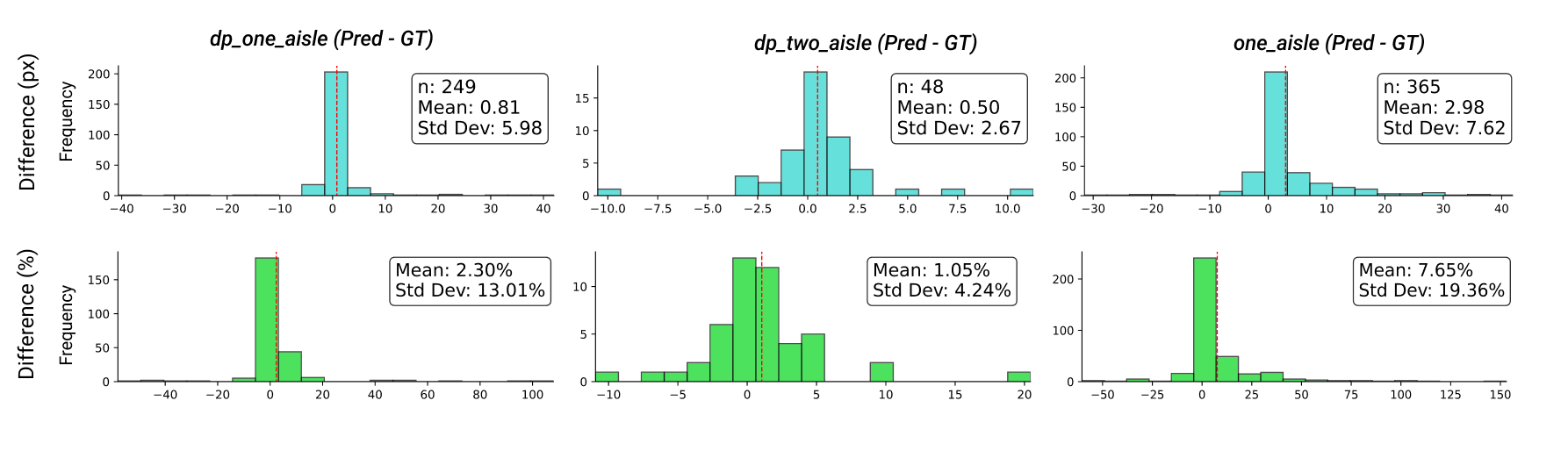}
    \caption{Results of the pipeline's derivation of width on the test set, for the classes with over 30 samples: \textit{dp\_one\_aisle}, \textit{dp\_two\_aisle}, and \textit{one\_aisle}. Other classes are not shown, as histogram patterns are less visible and robust with the less common classes. Differences in pixels above are shown in cyan, and percentage difference below in green. Differences are predicted minus ground truth; therefore, a positive value indicates overestimation. The model sees a large peak around zero (\textit{i.e.}, accurate estimation), with wider variation depending on class.}
    \Description{Histograms showing the model's width estimation error, for dp one aisle, dp two aisle, and one aisle classes. Both the percentage and difference in pixels is shown. All show a peak around zero, with spread highest for dp two aisle.}
    \label{fig:width-histograms}
\end{figure}

\subsection{Locator Models and Results}\label{locator_model_comparisons}
\begin{table*}[h]
\begin{tabular}{llll}
                    & \textbf{YOLOv11 large}                  & \textbf{DINO (ResNet50 backbone)}                                            & \textbf{Co-DETR (Swin-L backbone)}                                                              \\ \hline
\rowcolor[HTML]{F3F3F3} 
Training Epochs     & 200                                                                                                                                                   & 50                                                                  & 50                                                                                     \\
\rowcolor[HTML]{FFFFFF} 
Selected Epoch      & 126                                                                                                                                                   & 27                                                                  & 23                                                                                     \\
\rowcolor[HTML]{F3F3F3} 
Batch Size          & 16                                                                                                                                                    & 2                                                                   & 2                                                                                      \\
\rowcolor[HTML]{FFFFFF} 
Learning Rate       & $10^{-2}$                                                                                                                                             & $10^{-4}$                                                           & $2\times10^{-4}$                                                                       \\
\rowcolor[HTML]{F3F3F3} 
Weight Decay        & $5\times10^{-4}$                                                                                                                                      & $10^{-4}$                                                           & $10^{-4}$                                                                              \\
\rowcolor[HTML]{FFFFFF} 
Augmentations       & \begin{tabular}[c]{@{}l@{}}Hue\\  Saturation\\  Brightness\\  Translation\\  Resizing\\  Flipping (left right only)\\  Mosaic\\  Erasing\end{tabular} & \begin{tabular}[c]{@{}l@{}}Resizing\\  Random cropping\end{tabular} & \begin{tabular}[c]{@{}l@{}}Resizing\\  Random flipping\\  Random cropping\end{tabular} \\
\rowcolor[HTML]{F3F3F3} 
Precision/Recall/F1 & 0.76 / 0.92 / 0.83                                                                                                                                    & 0.87 / 0.86 / 0.86                                                  & 0.85 / \textbf{0.94 / 0.89}                                                           
\end{tabular}
\caption{Training parameters and results for the three models tested for the object detection component of the pipeline. Co-DETR sees the best performance in Recall and F1 score.}
\end{table*}

\end{document}